%% file: main.tex
\documentclass[sigconf]{acmart}
\usepackage{amsmath,amssymb,amsfonts}
\usepackage[ruled,vlined]{algorithm2e}
\usepackage{tabularx}
\usepackage{subfig}
\usepackage{graphicx}
\usepackage{physics}
\usepackage{textcomp}
\usepackage{xcolor,soul}
\usepackage{fancyhdr}
\usepackage{enumitem}
\usepackage{afterpage}
\usepackage{array}
\usepackage{bm}
\newcommand{\sol}{\textsc{Ribbon}}
\newcommand{\random}{\textsc{RANDOM}}
\newcommand{\hill}{\textsc{Hill-Climb}}
\newcommand{\rsm}{\textsc{RSM}}
\usepackage{tcolorbox}
\newtcolorbox{highlighted}{colback=yellow,coltext=black,breakable}
\definecolor{red_color}{rgb}{1.0, 0.0, 0.0}

\definecolor{blue_color}{rgb}{0.0, 0.0, 1.0}
\newcommand{\revision}[1]{\textcolor{black}{#1}}

\renewcommand\footnotetextcopyrightpermission[1]{}

\begin{document}

%\title{\sol{}: Serving Deep Recommendation Models using Heterogeneous Cloud Instances}

\title{\sol{}: Cost-Effective and QoS-Aware Deep Learning Model Inference using a Diverse Pool of Cloud Computing Instances}

\author{Baolin Li}
\affiliation{%
  \institution{Northeastern University}}
%   \city{Boston}
%   \state{MA}}
% \email{li.baol@northeastern.edu}

\author{Rohan Basu Roy}
\affiliation{%
  \institution{Northeastern University}}
%   \city{Boston}
%   \state{MA}}
% \email{basuroy.r@northeastern.edu}

\author{Tirthak Patel}
\affiliation{%
  \institution{Northeastern University}}
%   \city{Boston}
%   \state{MA}}
% \email{patel.ti@northeastern.edu}

\author{Vijay Gadepally}
\affiliation{%
  \institution{MIT Lincoln Laboratory}}
%   \city{Lexington}
%   \state{MA}}
% \email{vijayg@ll.mit.edu}

\author{Karen Gettings}
\affiliation{%
  \institution{MIT Lincoln Laboratory}}
%   \city{Lexington}
%   \state{MA}}
% \email{karen.gettings@ll.mit.edu}

\author{Devesh Tiwari}
\affiliation{%
  \institution{Northeastern University}}
%   \city{Boston}
%   \state{MA}}
% \email{d.tiwari@northeastern.edu}

\renewcommand{\shortauthors}{B. Li et al.}

\begin{abstract}

Deep learning model inference is a key service in many businesses and scientific discovery processes. This paper introduces \sol{}, a novel deep learning inference serving system that meets two competing objectives: quality-of-service (QoS) target and cost-effectiveness. The key idea behind \sol{} is to intelligently employ a diverse set of cloud computing instances (heterogeneous instances) to meet the QoS target and maximize cost savings. \sol{} devises a Bayesian Optimization-driven strategy that helps users build the optimal set of heterogeneous instances for their model inference service needs on cloud computing platforms -- and, \sol{} demonstrates its superiority over existing approaches of inference serving systems using homogeneous instance pools. \sol{} saves up to 16\% of the inference service cost for different learning models including emerging deep learning recommender system models and drug-discovery enabling models.

%Deep learning based personalized recommendation services are taking over the majority of compute cycles in the cloud due to their wide range of applications. In this paper, we performed an in-depth analysis of deep recommendation model inference serving in the cloud based on a complete characterization on AWS. With the goal of minimizing the inference serving cost while guaranteeing the quality of service (QoS), we propose a unique solution \sol{} to deploy a pool of heterogeneous cloud instances for real-time query serving. We designed a Bayesian-Optimization based scheme to assist cloud users in efficiently finding the optimal instance pool composition to deploy the model. We have shown that on real-world recommendation applications, the heterogeneous solution is capable of saving up to 14\% of the serving cost while satisfying the QoS compared to a homogeneous solution.

%by 8\% compared to the next best technique.

\end{abstract}
\settopmatter{printacmref=false}
\maketitle

\pagestyle{empty}

\input{./sections/introduction.tex}

\input{./sections/background.tex}

\input{./sections/motivation.tex}

\input{./sections/design.tex}

\input{./sections/methodology.tex}

\input{./sections/evaluation.tex}

%\input{./sections/related_work.tex}
\input{./sections/conclusion.tex}

\begin{acks}
% We thank anonymous reviewers for their constructive feedback. We are grateful to first-line responders and essential workers who have worked tirelessly to keep our community safe and functioning during the COVID-19 pandemic, and hence, enabled us to devote time to performing this research work. 
We acknowledge support from NSF awards 1920020, 2124897, and 1910601, MGHPCC, and Northeastern University. This research was also sponsored by the United States Air Force Research Laboratory and the United States Air Force Artificial Intelligence Accelerator under Cooperative Agreement Number FA8750-19-2-1000. The views and conclusions contained in this document are those of the authors and should not be interpreted as representing the official policies, either expressed or implied, of the United States Air Force or the U.S. Government. The U.S. Government is authorized to reproduce and distribute reprints for Government purposes notwithstanding any copyright notation herein.
\end{acks}

\bibliographystyle{unsrtnat}
\bibliography{refs}

\end{document}

%% file: sections/introduction.tex
\section{Introduction}
\label{sec:intro}

\subsection{Background and Problem Formulation}
\label{subsec:back}

Deep learning (DL) has been widely used in various areas of applications in recent years. The widely adopted models in areas including scientific computing~\cite{oh2020deep,schutt2018schnet,lusch2018deep}, computer vision~\cite{o2019deep,farhadi2018yolov3,szegedy2017inception} and personalized recommendation~\cite{gupta2020deeprecsys,argyriou2020microsoft,hsia2020cross,gupta2020architectural} areas show the significance of studying these models.  While model training focuses on the model accuracy and does not have strict time requirements, model inference is often conducted in real-time environments under \revision{quality-of-service (QoS)} constraints. In this work, we focus on the model inference, where the models are set up to serve a continuous stream of queries submitted by the model users.

%Although the architectural performance characteristics between different models can vary significantly~\cite{hsia2020cross,gupta2020architectural}, they still share a general goal during training and inference.

%Personalized recommendation services, used in a wide range of applications, are now consuming a large fraction of compute cycles in data centers~\cite{gupta2020architectural,argyriou2020microsoft}. A new type of learning-based models, referred to as deep recommendation models (DRMs), are driving these recommendation services in many data centers including Facebook and Alibaba~\cite{naumov2019deep,zhou2018deep,zhou2019deep}. DRMs are fundamentally different in their structure than traditional content-based filtering, collaborative filtering models or deep learning models~\cite{hsia2020cross,he2017neural}. The structure of deep recommendation models includes a combination of embedding tables and deep neural networks (DNNs) because these models need to process both sparse and dense types of input features. The embedding table structure is designed for efficient processing of sparse input features and DNNs are composed to efficiently process dense input features. 
%Consequently, as recent works have shown, these models exhibit different architectural performance characteristics and sensitivity toward different architecture resources than traditional deep learning models (e.g., QoS-constrained query inference for DRMs can be better on CPUs than GPUs in many cases)~\cite{hsia2020cross,gupta2020deeprecsys}. 

A few recent works have started to investigate and quantify the architectural performance bottleneck of various models when serving inference on different types of hardware architectures available in their data centers and platforms~\cite{jouppi2017datacenter,hsia2020cross,naumov2020deep,park2018deep,lane2015early}. 
\textit{However, these works do not address the cost-effectiveness of executing inference queries on different types of hardware. As DL models are becoming more widely adopted and used, inference service providers are more likely to utilize cloud computing resources to serve their customers. Thus, cost-effectiveness is becoming a primary optimization goal along with meeting the QoS targets.} Cloud computing platforms offer a wide range of processor architectures and hence, how to properly make use of the various available resources for DL models to meet the QoS targets and minimizing the cost becomes a key challenge. %In fact, our experiments reveal choosing the optimal architecture even for a single objective (cost or performance) is non-trivial for DRMs. 
This work addresses this challenge and advances the state-of-the-art by demonstrating how to serve inferences effectively under two competing objectives: QoS target and cost-effectiveness.
\subsection{Contributions}
\label{subsec:contribution}
This work makes the following contributions.

\textbf{\noindent{\newline{First, we provide evidence to demonstrate the performance and cost trade-off in serving deep learning inferences on various AWS cloud computing instances, and present a novel approach to exploit this trade-off.}}} Our experimental characterization reveals that the performance ranking of the different cloud computing instances can be often different than their cost-effectiveness ranking, even when considering only a single model. This characteristic makes it difficult to find the optimal computing instance for serving model inferences when co-optimizing for cost and QoS together. To exploit this trade-off between cost and QoS, this work approaches the problem with a new perspective: instead of exhaustively looking for a cheaper instance to serve the model with similar or better performance than the current one, it adopts the higher-performance and lower-cost instances to create a diverse pool to serve incoming queries cooperatively. 

%Our study shows that the trade-off between cost-effectiveness and performance exists among different cloud computing instances. We 

%Extensive results have shown that, if the diverse pool is configured properly, the cost can be significantly lowered.

%This work further advances our understanding by exploring the cost-effectiveness of DRM inferences. In particular, we show that simply provisioning more expensive computing instance does not lead to better performance -- in fact, it may be only marginally better than the average performance achieved via a wide range of different instance types. Furthermore, the performance ranking of the different cloud computing instances is often different than their cost-effectiveness ranking, even when considering only a single DRM. These characteristics make it difficult to find the optimal computing instances for serving DRM inferences when co-optimizing for cost and QoS together.

\textbf{\noindent{\newline{Second, this work introduces design and implementation of a novel framework, \sol{}~\footnote{\revision{\sol{} (\underline{R}equest \underline{I}nferencing \underline{B}ased on \underline{B}ayesian \underline{O}ptimizatio\underline{N}) binds together a set of diverse computing instances for serving model inference queries.}}, that minimizes the cost of serving a stream of inference queries while meeting the QoS target.}}} \textit{The key idea behind this framework is to build a pool of diverse computing instances (e.g., a combination of different CPU and GPU instance types) to serve model inferences instead of provisioning multiple instances of the same type (a traditional homogeneous pool).} Intuitively, \sol{}'s approach of building a diverse pool is rooted in the insights obtained from our experimental study: the optimal instance type changes depending on the model architecture, QoS target, and query batch sizes. Therefore, building a pool of instance types that are of high quality in different dimensions helps us achieve better outcomes when optimizing for multiple competing objectives (QoS target and cost). However, naively building a diverse pool of multiple computing instance types may lead to worse performance and cost. \sol{} formulates this as a search-space optimization problem and designs a Bayesian Optimization (BO)-based technique to find the optimal configuration from the diverse pool of multiple computing instance types.

\textbf{\noindent{\newline{Finally, \sol{}'s evaluation is based on inference streams emulating production-environment behavior~\cite{li2016work,gupta2020deeprecsys,gan2019open,hauswald2015sirius,kasture2016tailbench} and different types of DL models, such as CANDLE (predicts tumor \revision{cell} line response to drug pairs) and deep learning recommender system models~\cite{gupta2020deeprecsys,hsia2020cross,gupta2020architectural}.}}} Our evaluation confirms \sol's effectiveness over multiple competing strategies for multiple representative deep learning models on a range of AWS cloud computing instances. In particular, we show that \sol{} can increase the cost-effectiveness by up to 16\% while meeting the QoS target. \sol{} consistently outperforms competing techniques under a wide range of scenarios including different QoS targets, query characteristics, and load fluctuations. 

\textbf{Open-source artifact.} \sol{} framework is available at: \newline \url{https://doi.org/10.5281/zenodo.5262865}.

%\sol{}'s open-source contribution also opens up opportunities for further advancement in this emerging area (e.g., additional optimization objectives such as energy)

%and our open-source contribution will enable accelerating the advancement. \revision{\sol's software artifact is open-sourced for reproducibility and enables further improvements by the community.}

\subsection{Related Work}
\label{subsec:related_work}

%\revision{This section discusses relevant prior works. In particular, we compare and contrast \sol{} against existing works and discuss how the ideas in \sol{} advance our current understanding/capabilities.}

\revision{In this section, we compare \sol{} against prior works and discuss how the ideas in \sol{} advance current capabilities.}

Recent works on memory-heavy models (e.g., recommendation models) have provided insights on the new challenges such workload brings on both training and inference~\cite{gupta2020architectural,kalamkar2020optimizing,naumov2020deep}. DeepRecSys~\cite{gupta2020deeprecsys} builds a systematic approach to split and offload inference queries for higher production-scale throughput. However, none of the prior works have designed an online serving system in a multi-platform cloud environment setting, and exploited the diversity of compute instances for co-optimizing QoS target and cost-effectiveness (\revision{the core contributions of \sol}). 

Cloud providers offer a variety of configurations for cloud instances~\cite{yadwadkar2017selecting,cortez2017resource,kosta2012thinkair,li2018performance}. Traditionally, application developers provision instances based on expertise and tuning~\cite{rzadca2020autopilot,fontoura2019predictive,xu2017cred,dukic2019happiness,hsu2018micky,scheuner2018estimating}. However, with increasing application types and cloud servers, this approach is inefficient due to its production overhead~\cite{hwang2015cloud,li2010cloudcmp,scheuner2018cloud,shahrad2016availability}. This led to the development of application-specific tools for optimal instance selection with the aim to maximize performance~\cite{ferguson2012jockey,verma2011aria,venkataraman2016ernest,jalaparti2012bridging,shi2014mrtuner,jiang2016webperf,zhang2019mark}. These approaches often rely on detailed application profiling prior to execution~\cite{yadwadkar2017selecting,xu2013bobtail,ahn2012dynamic,moghaddam2018energy,baset2012towards,zhou2021mocha}. In some cases, they rely on historical training data from similar applications running on similar hardware for performance optimization~\cite{hsu2018arrow,mai2020kungfu,kundu2012modeling,wagenlander2020spotnik,jo2017machine,zhang2020model,chiang2014matrix,garcia2020cloud}. However, historical performance data is not available for the problem \sol{} is solving  and such data will lose significance with an update in the application or the hardware~\cite{wang2017learning,unger2020context,fang2020deep,kalamkar2020optimizing,naumov2020deep}. These approaches neither consider the vast cloud-based compute and accelerator options, nor the query stream of different batch sizes to models. %, which impacts the QoS targets~\cite{xu2012vslicer,psychas2017non,chung2018stratus,rao2014towards,rao2013optimizing}. 

Recent serverless inference frameworks such as MArk and BATCH ~\cite{zhang2019mark,ali2020batch} cannot be directly applied to recommendation models with large embedding tables and other general large-scale models since they would not fit in limited serverless memory space, and the serverless underlying hardware is not exposed to the user, making it difficult to exploit hardware diversity. Existing automatic cloud instance suggestion frameworks~\cite{alipourfard2017cherrypick,li2020automating} do not explore the idea of a diverse instance pool like \sol{}. Several scheduling frameworks spawn queries among cloud instances that scale up depending on the progress of the first few tasks~\cite{delimitrou2013paragon,delimitrou2014quasar, yeung2020towards,lu2017imbalance,ukidave2016mystic,park2012locality}. However, unlike \sol{}, they either do not deal with heterogeneous instances or assume that the queries will not change with time~\cite{amaral2017topology,asyabi2020akita}. \sol{} introduces a novel use case of Bayesian Optimization for building efficient inference service systems -- expanding the surface area of what we had demonstrated previously for other use cases of Bayesian Optimization in the data center/cloud computing resource-management area (i.e., shared resource partitioning for microservices~\cite{patel2020clite}, shared resource management for fairness and performance~\cite{roy2021satori}, and performance auto-tuning~\cite{roy2021bliss}). None of these previous works exploit the diversity in hardware instances for serving cost-effective model inference queries.

%% file: sections/background.tex
\section{Background}
\label{sec:backg}

\begin{table}[t]
\centering
\caption{Deep learning models used in this work.}
\vspace{-0.3cm}
\scalebox{0.83}{
\begin{tabular}{|p{0.10\textwidth}|p{0.40\textwidth}|} 
 \hline
 \textbf{Model} & \textbf{Description} \\ [0.5ex] 
 \hline
 CANDLE~\cite{xia2018predicting} & A large-scale fully-connected DNN model in Cancer Distributed Learning Environment (CANDLE) project~\cite{wozniak2018candle,candle}. Predicts tumor cell line response to drug pairs. \\ 
 \hline
 ResNet~\cite{he2016deep} & Developed by Microsoft, a CNN model with residual operations. Widely applied in computer vision areas such as image classification, object detection. \\
 \hline
 VGG~\cite{simonyan2014very} & A CNN model available on DLHUB~\cite{chard2019dlhub}. Widely applied in image recognition areas. \\
 \hline
 MT-WND~\cite{zhao2019recommending} & Multi-Task Wide and Deep, a recommendation model. Uses multiple DNN predictors in parallel to predict multiple metrics such as click-throughput rates (CTRs), ratings. Used for YouTube video recommendation. \\
 \hline 
 DIEN~\cite{zhou2019deep} & Deep Interest Evolution Network developed by Alibaba for recommendation. Integrates gated recurrent units (GRUs) to capture time series information. Used in e-commerce recommendation.\\
 \hline  
\end{tabular}}
\label{table:1}
\vspace{-0.4cm}
\end{table}

In this section, we provide a brief background on deep learning models and the cloud computing instances used in this study. 

%We also introduce a key characteristic of inference serving: the requests are served in batches and the batch size varies across queries.

\noindent{\textbf{\newline Studied inference models.}}To demonstrate the effectiveness and wide application range of \sol{}, the models in Table~\ref{table:1} cover a wide range of deep learning applications. A description for each model is also available in the table. These models can be grouped into two major categories: the general deep neural network (DNN), convolutional neural network (CNN) models as one, and the recommendation-specific DNN and embedding table hybrid models as another.

\begin{figure}[t]
    \centering
    \includegraphics[scale=0.15]{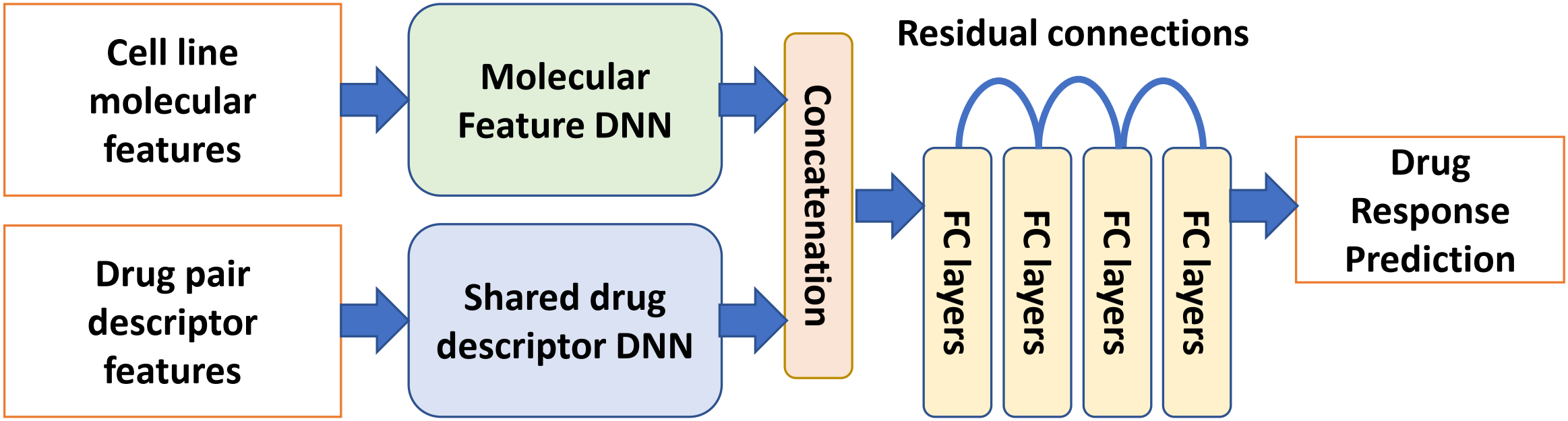}\newline
    %\vspace{0.1cm}
    \hrule
    \vspace{-0.3cm}
    \caption{CANDLE model architecture.}
    \label{fig:bkgd_candle}
    %\vspace{-0.5cm}
\end{figure}

The reason to include the CANDLE, ResNet, and VGG models is that they are very representative of the deep learning workloads in scientific research. CANDLE is a well-known cancer research project that applies deep learning techniques to biomedical studies, the model architecture is shown in Fig.~\ref{fig:bkgd_candle}. It combines the output feature of different DNNs to feed into a residual network for tumor's drug response prediction. The CANDLE model size is larger than other models as it combines multiple DNNs. The residual network is also the key feature in ResNet models used in our study. ResNet and VGG are widely studied models in computer science, specifically in computer vision applications. %Both of them are first-place winners of the ImageNet Large Scale Visual Recognition Challenges (ILSVRC)~\cite{russakovsky2015imagenet}. 

On the other side, MT-WND and DIEN are two representative workloads from industry personalized recommendation applications. Such applications are now consuming a large fraction of compute cycles in data centers according to the recent reports~\cite{gupta2020architectural,argyriou2020microsoft}. Therefore, studying and evaluating \sol{}'s effectiveness on such models is as important as the general DNN/CNNs. Current state-of-the-art models take a hybrid approach, which is fundamentally different in their structure and characteristics from traditional deep learning models~\cite{hsia2020cross}. Each input sample (e.g., a platform user and a product pair) consists of \textbf{categorical features} for the user's past interaction with products, as well as \textbf{continuous features} for user-specific information such as age.

\begin{figure}[t]
    \centering
    \includegraphics[scale=0.25]{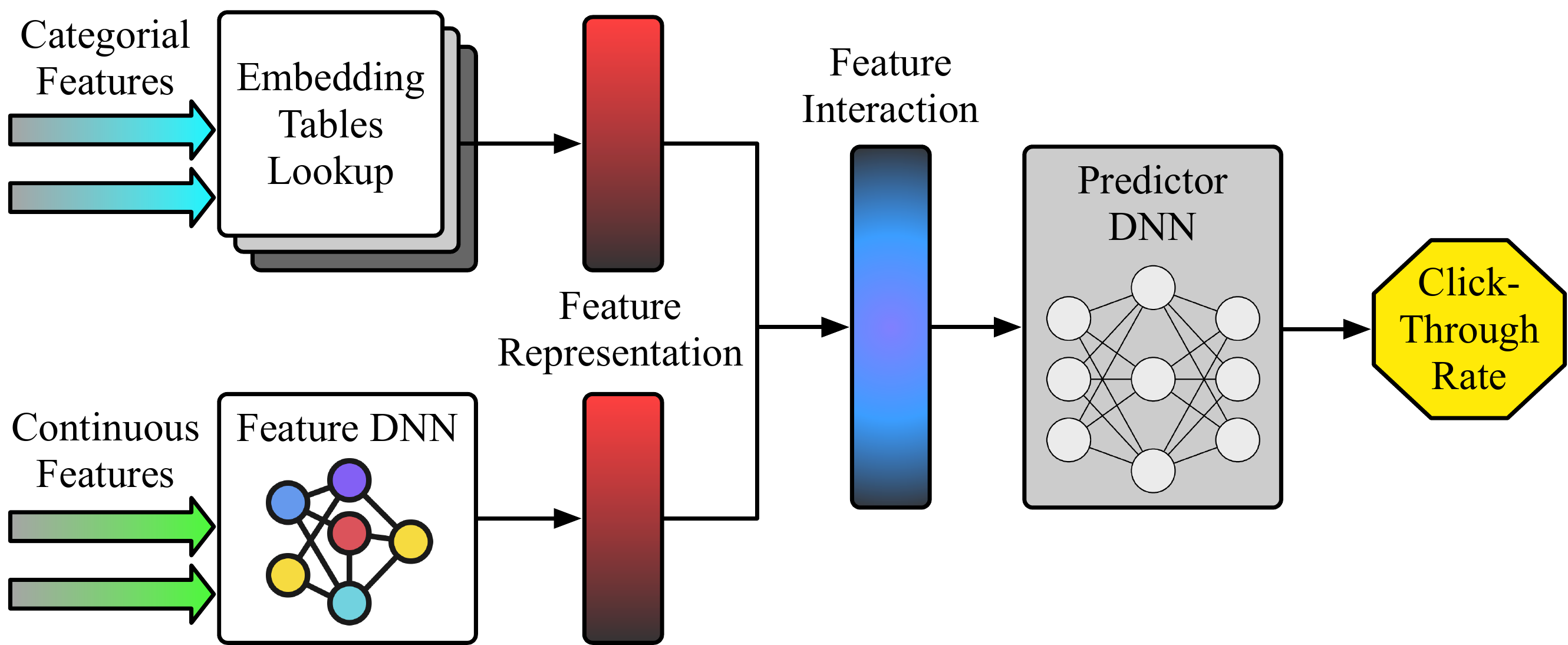}\newline
    %\vspace{0.1cm}
    \hrule
    \vspace{-0.3cm}
    \caption{Recommendation model architecture.}
    \label{fig:bkgd_1}
    %\vspace{-0.4cm}
\end{figure}

As shown in Fig.~\ref{fig:bkgd_1}, categorical features are processed by embedding tables and continuous features are processed by DNNs, their outputs get combined and fed to a final DNN for click-through rate (CTR, probability that user clicks on the product) prediction. The size of embedding tables is typically large due to the high number of product categories (in the order of tens of GBs of memory).

\begin{table}[t]
\centering
\caption{\revision{Studied AWS instances. A brief description with example application fields is provided for each category.}}
% \vspace{-0.2cm}
\scalebox{0.80}{
\begin{tabular}{|p{0.08\textwidth}|p{0.07\textwidth}|p{0.055\textwidth}|p{0.30\textwidth}|} 
 \hline
 \multicolumn{1}{|c|}{\textbf{Category}} & \multicolumn{1}{|c|}{\textbf{Instance}} & \multicolumn{1}{|c|}{\textbf{Size}} &
 \multicolumn{1}{|c|}{\textbf{\revision{Description}}} \\ [0.5ex] 
 \hline
 General purpose & \texttt{t3, m5, m5n} & \texttt{xlarge} & \revision{t3, m5, and m5n families offer a balance of compute, memory, and network resources (powered by different processor types). } \\ 
 \hline  
 Compute optimized & \texttt{c5, c5a} & \texttt{2xlarge} & \revision{c5 and c5a family instances are suitable for compute-intensive (c) workloads, but at higher price. c5a uses AMD EPYC and c5 uses Intel Cascade Lake processor.} \\ 
 \hline  
 Memory optimized & \texttt{r5, r5n} & \texttt{large} & \revision{r5 and r5n family instances are suited for memory-intensive workloads ('r' is a codename for memory-optimized).}\\  
 \hline  
 Accelerator (GPU) & \texttt{g4dn} & \texttt{xlarge} & \revision{g4dn family instances are GPU-based cost-effective instances -- targeted for machine-learning and graphic-intensive workloads ('g' refers to graphics/GPU).}\\  
 \hline
\end{tabular}}
\label{table:2}
% \vspace{-0.6cm}
\end{table}

\noindent{\textbf{\newline Inference Service in the Cloud.}} To efficiently serve the model inference queries under the QoS and budget constraints, we evaluate a wide variety of Amazon's AWS cloud computing EC2 instances. We chose AWS EC2 as a primary experimental playground because it provides a rich number of node instance types at different price points (e.g., compute-optimized, memory-optimized, accelerators, etc.). The example EC2 instances used throughout the paper are shown in Table~\ref{table:2}. 

%This work does not include any storage optimized instances as we find that storage performance does not impact inference task latency much. For the GPU instances, only \texttt{g4dn} instance (NVIDIA T4 GPU) is shown as we find it to have the best performance and lowest cost among all GPU instances. For large models like CANDLE, it cannot fit into the memory of the standard model sizes in Table~\ref{table:2}, thus the size for all instance types for this model needs to be scaled up by 2. 

One common characteristic of online inference service is that each query contains multiple requests, the number of requests in a query is called batch size. The batch size varies across queries to the model, and the reason is different for general deep learning models and the recommendation models. In the former case, since adaptive batching is widely applied in deep learning inference~\cite{crankshaw2017clipper,zhang2019mark,ali2020batch}, the number of requests that are batched into a query varies over time. In the latter case, since a query includes multiple products that need to be ranked and recommended to the user at the same time, it also varies between different queries~\cite{gupta2020architectural,gupta2020deeprecsys,hsia2020cross}. 

\noindent{\textbf{\newline Figure of Merit.}} An important metric for serving DL models in the cloud is the mean service latency and tail latency (more methodological details provided in Sec. ~\ref{sec:methodology}). We define the instance performance as its achievable throughput (queries per second, or QPS), which is the reciprocal of the mean service latency. Specifically, let the mean service latency be $L_{m}$, the QPS can be written as $\frac{1}{L_{m}}$. A higher performance indicates that the latency of the queries served on this instance is lower, thus more queries can be processed per unit time. When serving an online query stream, there is a Quality-of-Service (QoS) target as online service is latency-critical. The target is typically specified in terms of tail latencies. Another metric associated with serving is the price per instance, which is defined by the AWS cloud service provider.

%% file: sections/motivation.tex
\section{Opportunity and Challenges}
\label{sec:motiv}
In this section, first we describe the trade-off between performance and cost in cloud inference environment. Then, based on these trade-offs, we show the possibility of improving a system that is already performing well by introducing a diverse instance pool. Finally, we show the new challenges that come with a diverse pool and how a \sol{}-like solution can be applied for cost-effective and QoS-aware inference service on a diverse pool of instances.

\subsection{Performance and Cost Trade-off}
\label{sec:motiv_tradeoff}

\begin{figure}[t!]
    \centering
    \subfloat[][]{\includegraphics[scale=0.4]{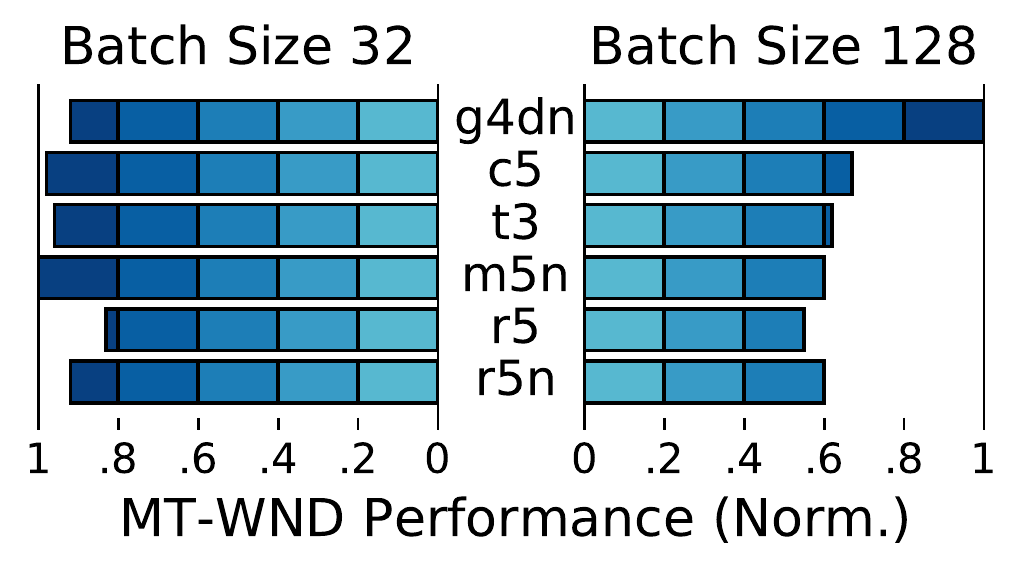}}\hfill
    \subfloat[][]{\includegraphics[scale=0.4]{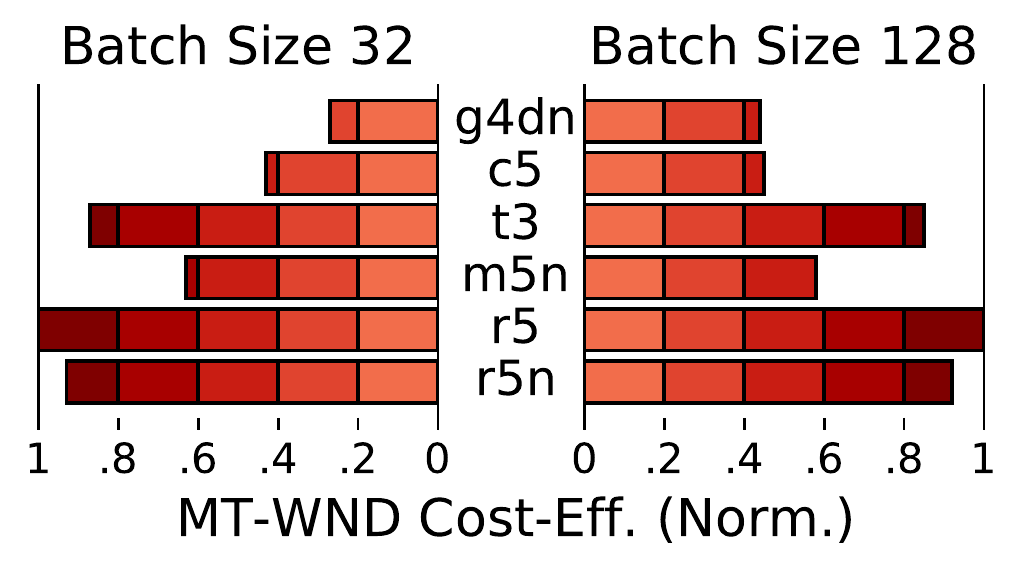}} \hfill\newline 
    %\vspace{0.1cm}
    \hrule
    \vspace{-0.3cm}
    \caption{Relative performance (a) and cost-effectiveness (b) of instances serving MT-WND model with \revision{batch} sizes 32 and 128. \revision{X-axis represents the normalized performance and cost-effectiveness as a score between 0 and 1. The normalization point is the best performance and most cost-effective instance, respectively. ``1'' indicates the best performance and the most cost-effectiveness, respectively.}}
    \label{fig:char_3}
    % \vspace{-0.4cm}
\end{figure}

%Recall from our discussion in Sec.~\ref{sec:backg}, multiple requests are grouped together into a batch with a batch size. 
Fig.~\ref{fig:char_3} (a) shows an example of how the batch size varies the relative performance of an instance type. For the MT-WND model with batch size 32, most instances shown have similarly high performance. But when the \revision{batch} size becomes 128, \texttt{g4dn} becomes the optimal instance, significantly outperforming other instance types. This is also an expected result as GPU's streaming multiprocessors are more suitable for inputs with larger batch sizes. 

Interestingly, the results in Fig.~\ref{fig:char_3} (b) tell a different side of the story about the cost-effectiveness. The cost-effectiveness is simply defined as the number of queries that an instance can process per dollar spent, as shown in the following Eq.~\ref{eq2}.
{\begin{align}
    \text{Cost\textendash Eff.} = \frac{\text{Perf. (query/sec)}}{\text{Price (\$/hour)}} = \frac{3600 \cdot \text{Perf.}}{\text{Price}} \text{ (query/\$)} \label{eq2}
\end{align}}
Even though the GPU instance \texttt{g4dn} has very strong performance for relatively larger batch sizes, its cost-effectiveness is the lowest among all instances shown. On the other side, the \texttt{r5} and \texttt{r5a} instances are low on the performance of higher batch sizes, but their cost-effectiveness are consistently the highest. This is because, with the same amount of memory allocated, the memory optimized instances incur less cost than the general purpose instances and much less than compute optimized and accelerator instances. Notice that the cost-effectiveness only relates to the number of queries that an instance can serve given a budget; it does not consider whether the service satisfies the latency requirement. If there is no latency requirement, it would be optimal to serve the models with \texttt{r5} or \texttt{r5n} instances. However, in a QoS-aware scenario, the cloud user would have to serve with \texttt{g4dn} just to stay within QoS requirements despite the extra cost. 

\subsection{\revision{Potential of Building Diverse Instance Pool}}

%Improving beyond Optimum}
\label{sec:motiv_improve}
For a deployed cloud service, we assume it is already running at minimal cost on a specific instance type - %among all available cloud instances types that can meet QoS of this service, the total cost of all currently deployed instances is the lowest.
reducing the number of total instances for lower cost will violate QoS. However, we ask ourselves a question: is it possible to further reduce the cost while not violating QoS? One common approach is to look for a more cost-effective instance type that also has the same or better performance so it does not violate QoS. But this is not always feasible as such instance type may not even exist. Instead, \sol{} takes an unorthodox approach -- it makes use of instance types that would violate the QoS with lower performance, but also with lower cost. Based on the discussions in Sec.~\ref{sec:motiv_tradeoff}, we can see that the cost-effective instance could still be useful for lower batch size queries even though it greatly violates the QoS for larger batches. If we form a diverse instance pool composed of both the higher performance cost-ineffective instances and the lower performance cost-effective instances, there is an opportunity that we can lower the total serving cost while staying within the QoS. 

\begin{figure}[t!]
    \centering
    \includegraphics[scale=0.55]{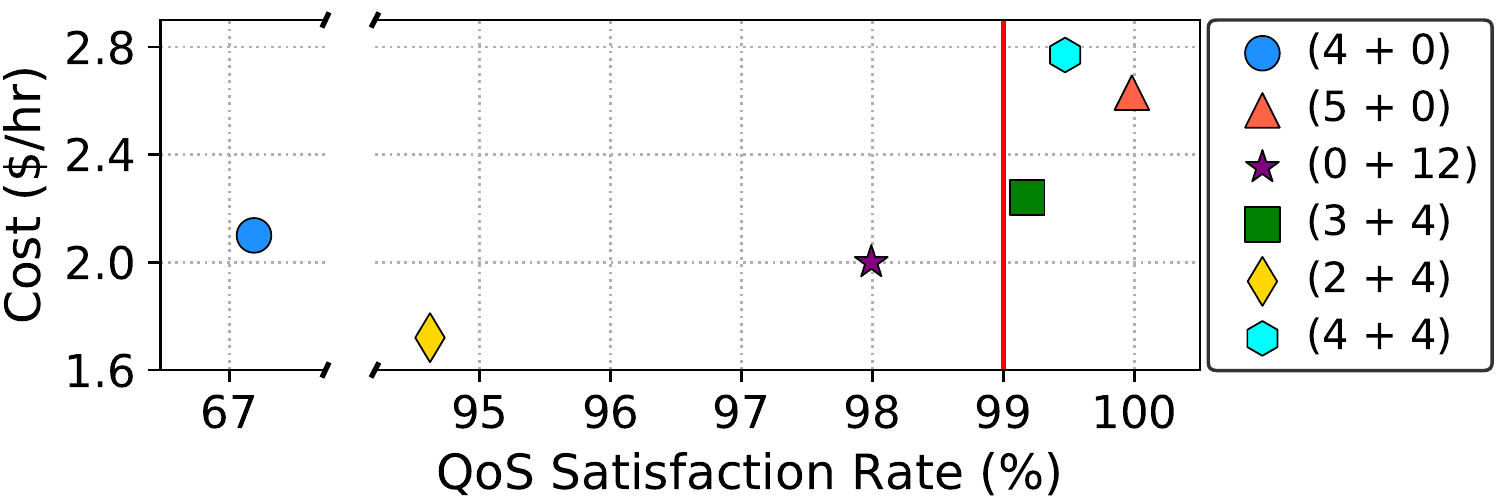}\newline
    %\vspace{0.1cm}
    \hrule
    \vspace{-0.3cm}
    \caption{MT-WND QoS satisfaction rate and service price for each instance configuration. The configuration (X + Y) means there are X \texttt{g4dn} instances and Y \texttt{t3} instances used. Configurations on the right of the red vertical line satisfy the QoS requirement.}
    \label{fig:char_7}
    % \vspace{-0.5cm}
\end{figure}

We illustrate such improvement opportunity using a straightforward example. Fig.~\ref{fig:char_7} quantifies the price and latency of serving a query stream of MT-WND model. The query inter-arrival times, batch size distributions and QoS target used in this example have been chosen to reflect production workloads~\cite{gan2019open,li2016work, gupta2020deeprecsys} (Sec.~\ref{sec:methodology}). The red vertical line indicates the QoS target, meaning at least 99\% of the queries have to be within the 99$^{th}$ percentile tail latency target (20ms). Among all instances, the \texttt{g4dn} instance is the optimal type to satisfy QoS with minimum cost, no other instance can replace \texttt{g4dn} for a lower cost. Meeting the QoS requires at least 5 \texttt{g4dn} instances in total (orange triangle), as only 4 such instances would significantly violate QoS (blue circle), thus 5 \texttt{g4dn} is the current optimum. 

To further improve the serving cost, we introduce the \texttt{t3} instance type. As shown in the plot (purple star), this instance is much cheaper with limited performance: 12 \texttt{t3} instance cannot satisfy the QoS, but the cost is still cheaper than 5 \texttt{g4dn} instances. If we use a diverse pool composed of both instance types, the serving cost can be further reduced while meeting the QoS target. The plot shows multiple heterogeneous or diverse pool configurations: (3+4), (2+4), and (4+4). We notice that the (4+4) configuration meets the QoS target, but is more expensive than the optimal homogeneous configuration (5+0). In fact, by reducing one instance of \texttt{g4dn} type to (3+4), our resulting configuration becomes less expensive than the (5+0) homogeneous pool configuration while meeting the QoS target. At this point, we have reached a lower serving cost by introducing the diverse pool serving. Further decreasing the \texttt{g4dn} instance number in the diverse pool to (2+4) will result in violation again, thus (3+4) becomes the new optimal configuration. The intuition behind diverse pool serving is that diversity increases our ability to balance the trade-offs (cost vs. performance) across query streams, instead of being one-dimensional. For example, cheaper instances can opportunistically help us reduce cost without violating the QoS when expensive instances are overloaded.

\subsection{Challenges in Building a Diverse Pool}
\label{sec:motiv_diversity}
While the diverse pool can further reduce serving costs beyond heterogeneous optimum, new challenges also appear with this new technique. The general guideline on picking instance types to add to the diverse pool is to use more cost-effective instances that can satisfy a relaxed QoS target. In the example in Fig.~\ref{fig:char_7}, we first relax the QoS target of 20ms for about 30\% to 26ms, and \texttt{t3} is an instance type that can satisfy the relaxed QoS while being more cost-effective than the \texttt{g4dn} as shown in Fig.~\ref{fig:char_3} (a). We find this degree of relaxation works well for all models during our experiments, instances selected with too much relaxation will not appear in the optimal diverse pool configuration as such instances would inevitably violate the QoS for its queries. As shown later in Sec.~\ref{sec:methodology}, this one-time profiling effort to form a diverse pool can also be skipped as the effective pool tend to be common for models of the same category. The real challenge is, once we have created a diverse pool, how to choose the number of each instance type such that the diverse configuration is optimal, i.e., in Fig.~\ref{fig:char_7}, how to quickly find the (3+4) configuration.

\begin{figure}[t!]
    \centering
    \includegraphics[scale=0.52]{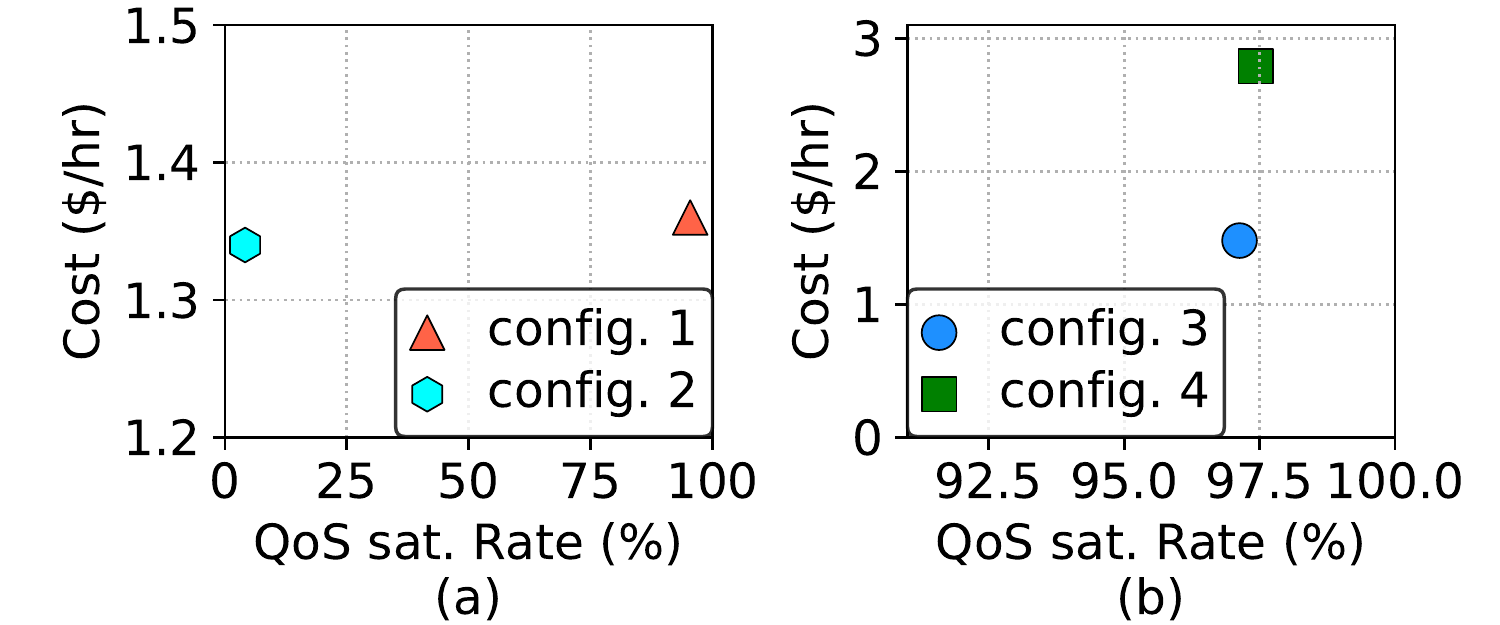}%\newline
    \vspace{1mm}
    \hrule
     \vspace{-3mm}
    \caption{(a) Configurations with similar cost but significantly different QoS rate. (b) Configurations with significantly different cost but similar QoS rate.}
    \vspace{-3mm}
    \label{fig:char_8}
\end{figure}

Finding the optimal diverse pool configuration is a non-trivial problem. Firstly, more dimensions are added to the search space as the number of each instance type represents a dimension. Due to higher dimensions, the search space becomes exponentially larger. Secondly, the interaction between the variable (a diverse pool configuration) and the corresponding QoS satisfaction rate cannot be described mathematically, resulting in ``costly'' evaluation for each configuration as the configuration has to be deployed to serve the queries to get its QoS satisfaction rate. This mathematically indescribable interaction is due to the unknown latency for each query and the randomness of query arrival and batch size. We assume \revision{no} prior/offline knowledge about the latency corresponding to queries with a specific batch size is available, therefore we cannot simply map queries of certain batch sizes to certain instance types. In this work, we apply a straight-forward first-come-first-serve policy for all queries (more details in Sec.~\ref{sec:methodology}). Thirdly, when searching for a better configuration, there is no clear sign to follow, and the result can be counter-intuitive. As shown in Fig.~\ref{fig:char_8} (a), configurations with similar cost may have significantly different QoS satisfaction rate, and in Fig.~\ref{fig:char_8} (b), configurations with significantly different cost (config. 4 has almost twice the price of config. 3) may still have similar QoS satisfaction rate. 

To address this challenge, \sol{} formulates this as an optimal configuration-finding optimization problem and designs a BO-based solution to find the optimal configuration with a minimum number of configuration evaluations.

%% file: sections/design.tex
\section{\sol{}: The Solution}
\label{sec:design}

\sol{} determines the optimal heterogeneous/diverse pool configuration via search space exploration process. At the start of the exploration process, it takes the instance types and QoS target as the inputs. The goal is to determine how many instances of each instance type should be chosen (the optimal pool configuration) such that it meets the QoS targets with minimum cost. The most challenging part is \textit{quickly} determining the number of instances for each type that leads to the optimal pool configuration -- the number of evaluations for different such configurations should be small. \sol{} uses Bayesian Optimization (BO) at its core to mitigate this challenge. BO is chosen because it is a lightweight online learning model that does not require expensive training, apriori knowledge, or data. Our evaluation demonstrates that \sol's BO is effective when compared against other competing search space exploration methods (Sec.~\ref{sec:methodology}).

\begin{figure}[t]
    \centering
    \includegraphics[scale=0.29]{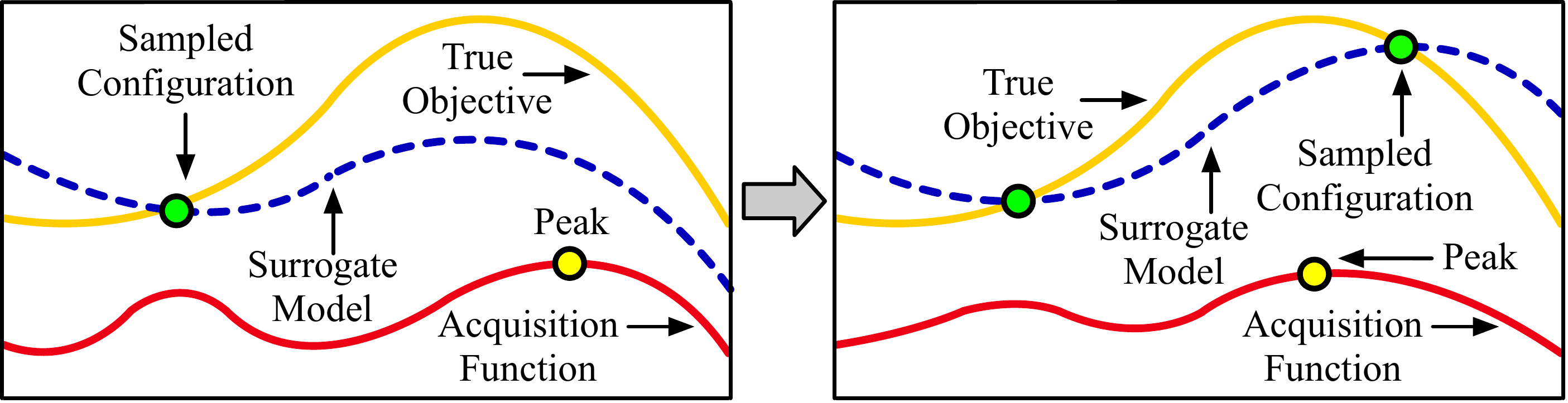}\newline
    %\vspace{0.1cm}
    \hrule
    \vspace{-3mm}
    \caption{The Bayesian Optimization approach used by \sol.}
    \label{fig:desi_1}
\end{figure}

\noindent{\textbf{\newline Working principles of \sol{}'s BO Engine.}} BO is a black-box optimization method for minimizing an unknown objective function ($f(x)$) and is typically used when the sampling of the configurations is costly~\cite{shahriari2015taking,osborne2009gaussian} -- \sol's context shares this similarity. Next, we briefly describe how \sol{} uses a BO-based decision making engine at its core (Fig.~\ref{fig:desi_1}).

Initially, \sol's BO engine has no knowledge of the nature and distribution of the unknown objective function ($f(x)$). BO maintains a \text{prior belief} of $f(x)$ through its \text{surrogate model}. This surrogate model serves as a proxy for the true objective function. The success of BO depends on how the surrogate model represents the true objective as the optimization proceeds. \sol{} selects Gaussian Process (GP) as its surrogate model since it is adaptive toward representing any random process and is not input- or objective-specific. This enables \sol{} to handle changing streams of requests over various cloud instances, which is critical in \sol's context. Other surrogate model choices such as Tree Parzen estimators and polynomial estimators are not suitable for \sol{} because they assume conditions that are not true for \sol{} (e.g., polynomial objective function, samples can be divided into classes).

As \sol{} starts sampling configurations, it gathers the true value of $f(x)$ at those configurations (\text{posterior}). \sol{} uses this knowledge to update the value of the surrogate model at sampled configurations. The surrogate model uses its \text{covariance kernel} to extrapolate the value of $f(x)$ for configurations that are not sampled yet. This extrapolated value is a function of two factors: (1) the covariance of sampled configurations, and (2) their distance from the not-yet-sampled configurations. \sol{} chooses the \revision{Matern 5/2 covariance kernel~\cite{rasmussen2003gaussian,snoek2012practical}} for ensuring smoothness, and it also provides the capability that similar configurations will result in similar objective values, a desirable property in the context of \sol{}. Other possible choices of covariance kernels include Dot Product and Rational Quadratic assume the configurations to follow a monotonic or a particular kind of polynomial distribution of objective values -- therefore, unsuitable for \sol.  

The surrogate model of \sol{} gets updated after each sample is evaluated. The surrogate model associates a confidence value with configurations that are not sampled. It is inversely proportional to the variance of $f(x)$ for the nearby sampled configurations. \sol{} needs to intelligently sample to move toward the optimal configuration. \textit{So, how does \sol{} decide what to sample next?} 

BO has an associated \textit{acquisition function} ($a(x)$) which helps \sol{} determine what to sample next. The acquisition function takes into account the surrogate model and the confidence of the surrogate model for configurations that are not sampled yet (higher confidence means more likelihood of the surrogate model to represent the true objective function) to decide the next best configuration to sample. \sol{} uses Expected Improvement (EI) as its acquisition function. \revision{For each unexplored configuration, EI uses its GP mean and variance as input and calculates the expected improvement over the best explored configuration. By maximizing this expected improvement over all unexplored configurations, BO aims to sample a new configuration that provides the highest improvement over the current best.}

This acquisition function is used by \sol{} to steer its way to the optimal configuration by actually sampling only a few configurations. \sol{} performs its optimization based on the exploration vs. exploitation of configuration space (exploration: exploring configurations with high uncertainty to find the value of the optimal configuration, and exploitation: exploring the vicinity of configurations where the objective function is already sampled). The acquisition function needs to maintain a balance between the two for the BO to effectively navigate through the configurations without losing the opportunity to sample better-performing ones. Since the optimization performance of \sol{} depends on the objective function $f(x)$, it is important to carefully design the objective function so that it represents the metric that \sol{} is optimizing. 

\noindent\newline\textbf{How does \sol{} design an effective objective function?} Recall that \sol{} jointly optimizes for two competing objectives: meeting the QoS target and minimizing the cost. Since these objectives are competing, it is critical to construct an objective that carefully captures \sol's goal: minimizing the cost while meeting the QoS target. Unfortunately, this objective cannot be expressed as a typical optimization function that can be minimized or maximized, unlike traditional use cases of BO~\cite{wada2019bayesian,NEURIPS2019_a7b7e4b2}. To address this challenge, \sol{} constructs a new objective function. 

\sol's objective function is guided by the following principle: \textit{when a configuration satisfies the query QoS, the objective function should guide the evaluation toward lower cost; when the diverse configuration does not satisfy the query QoS, the objective function should guide the evaluation towards a higher QoS satisfaction percentage}. More formally, suppose the input variable over $n$ instance types is $x=[x_1, x_2, ..., x_n]$ where $x_i$ indicates number of instances of type $i$, Eq. ~\ref{eq5} shows the objective function designed over the two optimization aspects of \sol{}:

{\begin{align}
    f(\bm{x}) = \begin{cases}
        \frac{1}{2} \cdot \frac{R_{sat}(\bm{x})}{T_{qos}} & \text{if violates QoS,} \\
        \\
        \frac{1}{2} + \frac{1}{2} \cdot (1-\frac{\sum_{i=1}^{n} {p_i \cdot x_i}}{\sum_{i=1}^{n} {p_i \cdot m_i}}) & \text{otherwise. } \\
        \end{cases}
        \label{eq5}
\end{align}}

$R_{sat}(\bm{x})$ is the QoS satisfaction rate achieved by the chosen configuration, constant $T_{qos}$ is the QoS target (i.e., 99\% of queries meet a latency target). Constants $p_i$ and $m_i$ are the unit time price and the upper bound of the number of instances for instance type $i$, respectively. This objective function returns a normalized output between 0 and 1, its design ensures that any configuration that satisfies the QoS is superior than a QoS violation configuration regardless of the serving price, as $0 \leq R_{sat}(\bm{x}) < T_{qos}$. Maximizing the objective function guarantees the minimization of the serving cost while meeting QoS. Note that while this objective function can be evaluated at specific instance configurations when a model is run on that configuration, its general shape over the entire configuration space is unknown ($R_{sat}(x)$ cannot be known without evaluating configuration $x$). Therefore, \sol{}'s BO attempts to obtain the true shape of the objective function and find its maximum with as few configuration evaluations as possible. 

\revision{The parameter, $m_i$,  determines the upper bound on the search range. $m_i$ corresponds to the maximum number of instances of a given type such that adding any more number of instances of the same type does not improve the QoS satisfaction rate. For example, when serving with $u$ instances of a certain type, the QoS satisfaction rate is 95\%, and this rate stays at 95\% even when serving with $u+1$ instances, then $m_i$ for this type is chosen as $u$.}

%\revision{
We also experimented with other objective functions (e.g., a traditional single-metric function that returns the total cost but only for configurations that satisfy QoS and minimize the returned cost), but such design did not work well. The major lesson learned from this exploration was that the objective function needs to consider both the criterion (cost and QoS) together. Secondly, the objective function needs to maintain smoothness. For a non-smooth single-metric objective function, a large portion of the search space will be flat, which cannot provide guidance for configurations with a better QoS satisfaction rate. This is why \sol{} designed Eq.~\ref{eq5} to increase smoothness over both QoS-satisfaction and QoS-violating regions. Furthermore, we learned that the boundary between the QoS-violating region and QoS-satisfaction region can make it difficult to optimize for the acquisition function over a sudden change, and hence a steep jump needs to be avoided. 

We also considered batch size distribution in the objective function, but we noted that changing the batch size only changes the search space, the underlying BO process will still find the optimal configuration. Also, we noted that exploiting batch size distribution information risks making the optimization benefits sensitive to such parameters, which may not be always desirable.
%Note that the objective function can be potentially further improved by incorporating the information about batch size. However, in our experience, the improvements were marginal at the expense somewhat increased exploration time due to additional modeling complexity. We also note that exploiting batch size distribution information risks making the optimization benefits sensitive to such parameters which may not be always desirable.}

%\revision{We note that \todo{FIXME} clarify why the order is based on increasing number of instances (sum of xi) instead of the sum of pi times xi (numerator from equation 2).}

Next, we discuss several other \sol{}-specific challenges that \sol{} needs to mitigate to make the search process more efficient. 

\noindent\newline\textbf{\sol{} maintains a smooth distribution of configurations.} Recall that BO considers the outcomes of neighboring sampled configurations to determine the next configuration to sample. Specifically, BO's underlying GP sets the surrogate model value of the not-yet-sampled configurations based on the variance of already-sampled neighboring configurations. Consequently, the surrogate model indirectly guides the acquisition function toward the optimal configuration. Therefore, it is important that the true objective values of neighboring configurations follow a smooth distribution (i.e., less variance in the objection function values). High variance makes the acquisition function repeatedly sample configurations in the vicinity of that neighborhood, losing the opportunity to explore potentially better configurations in other neighborhoods. 

\sol{} employs two specific measures to solve this problem. First, we note that QoS-violating configurations are sub-optimal regardless of how close they are in meeting the QoS, but \sol{} deliberately considers their performance in the objective function (Eq.~\ref{eq5}).  This makes the objective function smoother in neighborhoods with QoS-violating configurations. In a naive design, where all QoS violating configurations simply return 0, the objective function will have a large sudden change in value between QoS violating and QoS satisfying configurations -- \sol{} avoids this. (2) In individual dimensions (particular cloud instance type), \sol{} arranges the configurations in increasing order (increasing number of instances) before the BO begins to search. This guarantees a smooth distribution of configurations along each of the dimensions.
 
\begin{figure}[t!]
    \centering
    \includegraphics[scale=0.33]{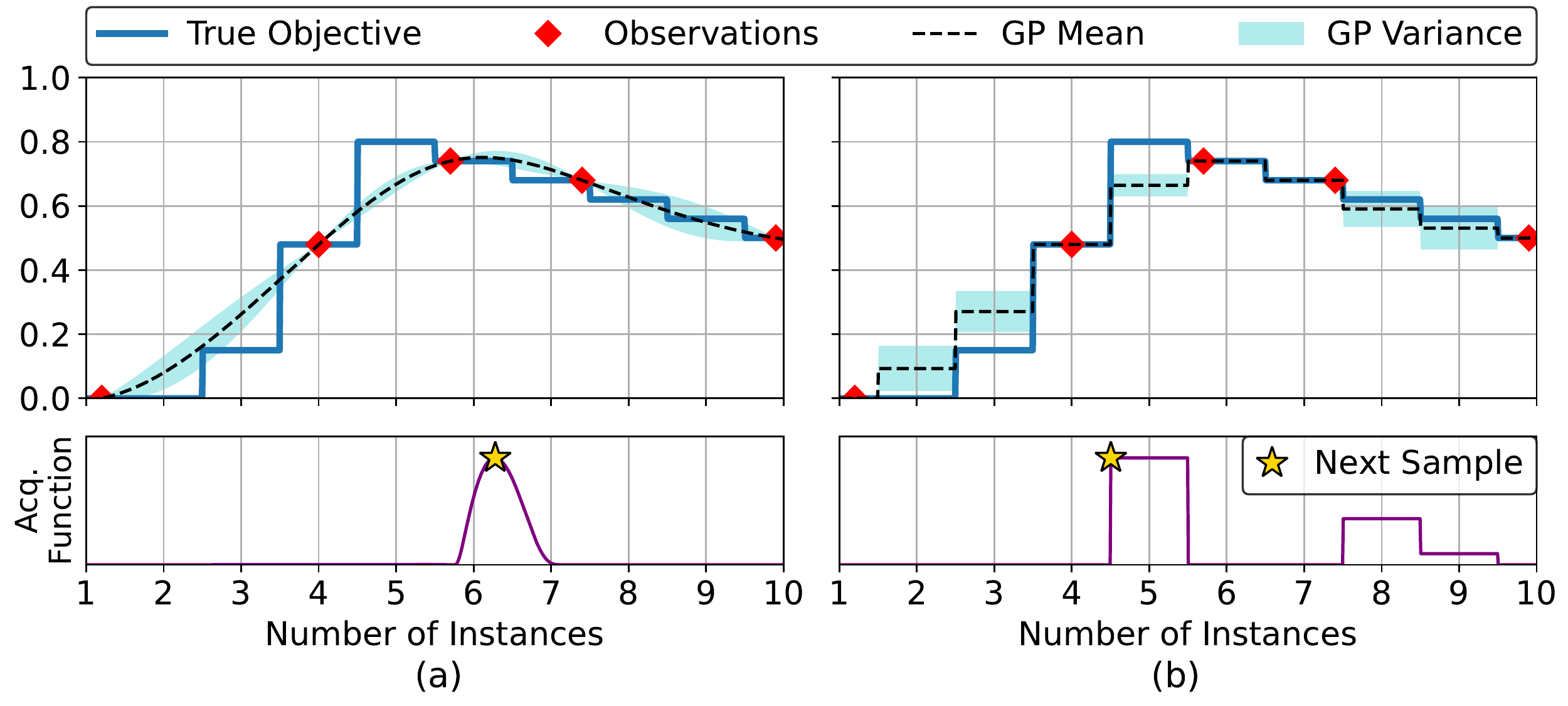}%\newline
    \hrule
    \vspace{-3mm}
    \caption{Effect of the rounding mechanism when varying the number of instances. For easier visualization, only one instance type is varied. The default BO behavior is shown in (a), and \sol's rounding off is shown in (b).}
    \label{fig:desi_3}
\end{figure}

\noindent\newline\textbf{\sol{} intelligently handles categorical variables.} By default, BO is designed to handle variables that have continuous values. However, in the case of \sol{}, BO samples configurations that only have categorical values as the number of cloud instances of each type are integers ~\cite{garrido2020dealing}. To tackle this issue, \sol{} applies a rounding technique to the underlying continuous GP kernel $k(x_i,x_j)$ of BO to form a rounded off GP kernel $k'(x_i,x_j)$:
{\begin{align}
    k'(\bm{x_i},\bm{x_j}) = k(R(\bm{x_i}), R(\bm{x_j})) \label{eq3}
\end{align}}
Here, $R(\bm{x})$ is the rounding function that rounds the variable $\bm{x}$ to its nearest integer. This ensures that the shape of the GP better matches the shape of the true objective function. Fig.~\ref{fig:desi_3}(a) is the default BO where the true objective has a step-like behavior due to the integer categorical values of the configurations. However, the GP (determined by the GP mean) has continuous behavior. In this case the GP does not represent the true objective well. Fig.~\ref{fig:desi_3}(b) shows how \sol{} uses the rounding mechanism on the GP, making the GP mean better represent the true objective function. The acquisition function remains consistent within the integer rounding range. In Fig.~\ref{fig:desi_3}(a), when the next sampled configuration falls within the integer range of previously sampled configurations, the acquisition function cannot provide any useful information. In contrast, with \sol{} (Fig.~\ref{fig:desi_3}(b)), the rounding mechanism guarantees the next sampled configuration to not fall into the same integer range of a previously sampled configuration. Therefore, it moves faster toward finding the next sample that optimizes the objective.

\revision{The above two techniques (objective function design and categorical parameter handling) make the optimization process feasible and convergent. For example, if we use a non-smooth objective function, BO’s acquisition function optimizer fails in 35\% of the cases and returns invalid values. Similarly, if \sol{} drops its mechanism to handle categorical parameters carefully, it would perform repetitive sampling of configurations that fall into the same integer range of previous sampled configurations, resulting in even more total samples than exhaustive search in more than 30\% of the cases. Using \sol{}’s design, we converged quickly for all our cases.}

\noindent\newline\textbf{\sol{} performs active pruning to speed up the search.} \sol{} uses its construction of objective function to prune the search space beyond what a default BO can do, further reducing the number of samples needed to reach the optimal configuration. 

Recall that as per \sol's objective function, a QoS-meeting configuration will be judged to be sub-optimal than the previously evaluated QoS-meeting configuration if the price is higher. A higher price configuration can also violate the QoS, hence, returning a value smaller than $\frac{1}{2}$, which is sub-optimal. \sol{} actively maintains a prune set $\mathbb{P}$ of such configurations that are unlikely to outperform the current optimal configuration. When a configuration $\bm{x_c} = [c_1, c_2, ..., c_n]$ is evaluated to violate the QoS by more than a threshold $\theta$ (e.g. 1\%), any configuration $\bm{x_c'} = [c_1', c_2', ..., c_n']$, where $\forall \text{ integer } 0 \leq i \leq n, c_i' \leq c_i$ cannot meet the QoS. Such configurations are added to the set $\mathbb{P}$. The set is updated after each sampling and is applied to the acquisition function as a constraint such that all the configurations in this set are avoided in the subsequent sampling. Whenever the acquisition function has the highest value for a configuration lying inside the set $\mathbb{P}$, \sol{}'s BO avoids sampling it and samples the next best configuration that has the highest acquisition function value among the ones not in $\mathbb{P}$. %Here \sol{}'s BO avoids sampling in the pruned region. 
As the optimization process progresses, \sol{}'s BO identifies more configurations that belong to $\mathbb{P}$. This decreases the number of configurations to search among, and hence, the speed towards the optimal solution increases. %Note that as \sol{} progresses, it prunes away configurations that may perform sub-optimally compared to the already sampled configurations. In this way, \sol{} attempts to search more efficiently as it navigates toward the optimal configuration. 

\noindent\newline\textbf{\sol{} promptly responds to load \revision{changes}.} Online service can experience fluctuations in the load. When such changes occur, \sol{} performs scaling at the instance level. \revision{\sol{} detects the load changes via performance monitoring  -- when the load goes up, more queries get queued in the query queue, and the QoS satisfaction rate will drop significantly due to the wait time. By monitoring the query queue size and the current QoS rate, one can determine whether the load has changed.}

While previous works~\cite{zhang2019mark,ali2020batch} have resort to serverless functions as a means of scaling, we find that firstly, serverless functions do not guarantee the latency target for a query as the user has no information about the underlying hardware the serverless function runs on. Secondly, some workloads such as the recommendation models come with large embedding tables requiring tens of GBs of memory, general deep learning workloads can also have a large model size and large data size, which easily exceed the memory limit of serverless functions. Therefore, \sol{} performs scaling at the instance level. If the serverless function can guarantee latency targets and the workload can fit into its memory, \sol{} is still compatible with serverless as users can use \sol{} for base instance scaling and use serverless functions to deal with strong bursty workloads. 

\sol{} maintains a complete record of the explored configurations in the instance space during its convergence to the optimal diverse pool. This information becomes useful when the load scales up as \sol{} will rely on the exploration record to avoid unnecessary sampling of low-QoS configurations and explore around the QoS satisfaction regions. After the scaling, the previous optimal configuration no longer satisfies the QoS target. The BO needs to restart the whole search process and find the new optimal configuration for the new load. A simple approach is to forget about the previous exploration results and restart BO from scratch. \sol{} tries to make use of the previous exploration results to help with the new BO process instead. 

In the beginning, the only explored point is the previous optimal configuration on the new load. Firstly, from the previous exploration record, \sol{} identifies the configurations that return the same or lower QoS satisfaction rate than the previous optimal. Let such configurations form a set $\mathbb{S}$. The intuition for collecting this set is that if the previous optimal cannot satisfy the QoS of the new load, any configuration that works as good as or worse cannot satisfy the new load QoS either. For each configuration in set $\mathbb{S}$, \sol{} performs the same operation introduced in pruning, and add all other configurations associated with this configuration to the pruning set $\mathbb{P}$. 

In addition, \sol{} estimates the objective function values of the configurations in $\mathbb{S}$ to feed to the new load BO as training data. Normally, the objective function value of an evaluation is only known after it has been evaluated by the BO process. However, since we know the configurations in $\mathbb{S}$ cannot meet QoS for the new load just as the previous optimal, we can estimate how much it violates the QoS. From our experiment, we find that the estimation does not need to be highly accurate, as its only purpose is to convey a message to BO, which tells the BO to not evaluate configurations near the set $\mathbb{S}$ configurations. We use a simple linear function to estimate the QoS satisfaction rate. For example, let the previous optimal configuration be A and another configuration in set $\mathbb{S}$ as B. If A has 99.9\% satisfaction rate on previous load and 33.3\% rate on new load, and B has 90\% satisfaction rate on the previous load, we estimate that B has 30\% satisfaction rate on new load. With both pruning and estimation applied based on exploration record on the previous load, the new BO process receives a head start. The related scaling results are shown in Sec.~\ref{sec:methodology}.

%% file: sections/methodology.tex
\section{Methodology and Evaluation}
\label{sec:methodology}

In this section, we analyze our evaluation results to understand different aspects of \sol{}, accompanied by further relevant methodological details corresponding to different evaluation focus. %\vspace{0.2cm}

%\todo{I will edit this part tomorrow, but a few requests: justify the QoS targets: it say within the range: can we be more specific?, citation for Gaussian distribution batch size distribution, why is it representative, can we try one more distribution? I removed the batch size distribution plots. can we try some other inter-arrival: although this is very well substantiated. multiple queries are served currently by the available pool of instances}

%\noindent\textbf{Evaluated workloads and query characteristics.} 

\subsection{Evaluated workloads} \vspace{1mm}

\sol{} is evaluated using the five models introduced in Sec.~\ref{sec:backg}: CANDLE, ResNet50, VGG19, MT-WND, and DIEN. These models cover a wide range of DL networks: DNN, CNN, embedding tables, RNN. 
\revision{An important experimental methodology detail is the query arrival and batch size distribution. Our evaluation is driven by an actual trace~\cite{gupta2020deeprecsys} and follows the same batch size and arrival distribution as other recent works in the area which have shown these characteristics are representative of the model inferences in production workloads ~\cite{li2016work, kasture2016tailbench,hauswald2015sirius,gan2019open}. Specifically, our batch size distribution follows a heavy-tail log-normal distribution, similar to previous work~\cite{gupta2020deeprecsys} which confirms that heavy-tail log-normal is more representative behavior than traditional log-normal distribution. We acknowledge that the batch size distribution for CANDLE can be potentially different from other models. Therefore, for further robustness of \sol{}, we evaluated \sol{} using Gaussian distribution-based batch size because Gaussian distribution can also capture some production workload behavior~\cite{li2016work}. Later in our evaluation and analysis, our results show that \sol{} benefits are not sensitive to the choice of these parameters.}

The query inter-arrival time follows a Poisson distribution, same as other works in this area~\cite{gupta2020architectural,li2016work,hauswald2015sirius,gan2019open,reddi2020mlperf,kasture2016tailbench}. Tail latency is a typical QoS target used in cloud service applications, in this paper we use the $99^{th}$ percentile tail latency as default. The query tail latency targets are selected within application Service Level Agreement (SLA) range, \revision{and ensured that the highest-performance instance (\texttt{g4dn} in our case) can satisfy the targets (otherwise no configuration can meet QoS). For the recommendation models, Google MT-WND and Alibaba DIEN both require an SLA target of tens of milliseconds~\cite{cheng2016wide,zhou2018deep}. Therefore, we set the MT-WND and DIEN QoS targets to be 20ms and 30ms, respectively --  similar to targets chosen by previous works for these models~\cite{gupta2020deeprecsys}. For other models, we set the QoS target to be within 1000ms, consistent with other works  ~\cite{li2019edge,gujarati2017swayam}. According to the latency for the largest query on the \texttt{g4dn} instance, we set the QoS targets for CANDLE, ResNet50, and VGG19 to 40ms, 400ms, and 800ms, respectively. These targets cover a wide range to demonstrate the effectiveness of \sol{} over a variety of latency targets. We confirmed that \sol's benefits are not sensitive to the chosen target QoS latency and achieve similar results when the QoS targets are varied.}

The query processing follows a simple first-come-first-serve (FCFS) manner, with the first arrived query going to the first available instance following the heterogeneous type order in Table~\ref{table:3}. \revision{We note that multiple queries are served concurrently by the available pool of instances.}

%For the GPU instances, only \texttt{g4dn} instance (NVIDIA T4 GPU) is shown as we find it to have the best performance and lowest cost among all GPU instances. For large models like CANDLE, it cannot fit into the memory of the standard model sizes in Table~\ref{table:2}, thus the size for all instance types for this model needs to be scaled up by 2. 

\begin{table}[t]
\centering
\caption{Cloud Computing Instance types used for different models (refer to Table~\ref{table:2} for AWS cloud instance code names).}
\scalebox{0.90}{
\begin{tabular}{|p{1.5cm}|p{2.8cm}|p{2.3cm}|} 
 \hline
 \textbf{Model} & \textbf{Homogeneous Pool} & \textbf{Diverse Pool} \\ [0.5ex] 
 \hline
 CANDLE & \texttt{c5a} & \texttt{c5a, m5, t3} \\ 
 \hline  
 ResNet50 & \texttt{c5a} & \texttt{c5a, m5, t3}\\ 
 \hline  
 VGG19 & \texttt{c5a} & \texttt{c5a, m5, t3}\\  
 \hline  
 MT-WND & \texttt{g4dn} & \texttt{g4dn, c5, r5n}\\  
 \hline
 DIEN & \texttt{g4dn} & \texttt{g4dn, c5, r5n}\\  
 \hline
\end{tabular}}
\label{table:3}
\end{table}

\begin{figure}[t!]
    \centering
    \includegraphics[scale=0.41]{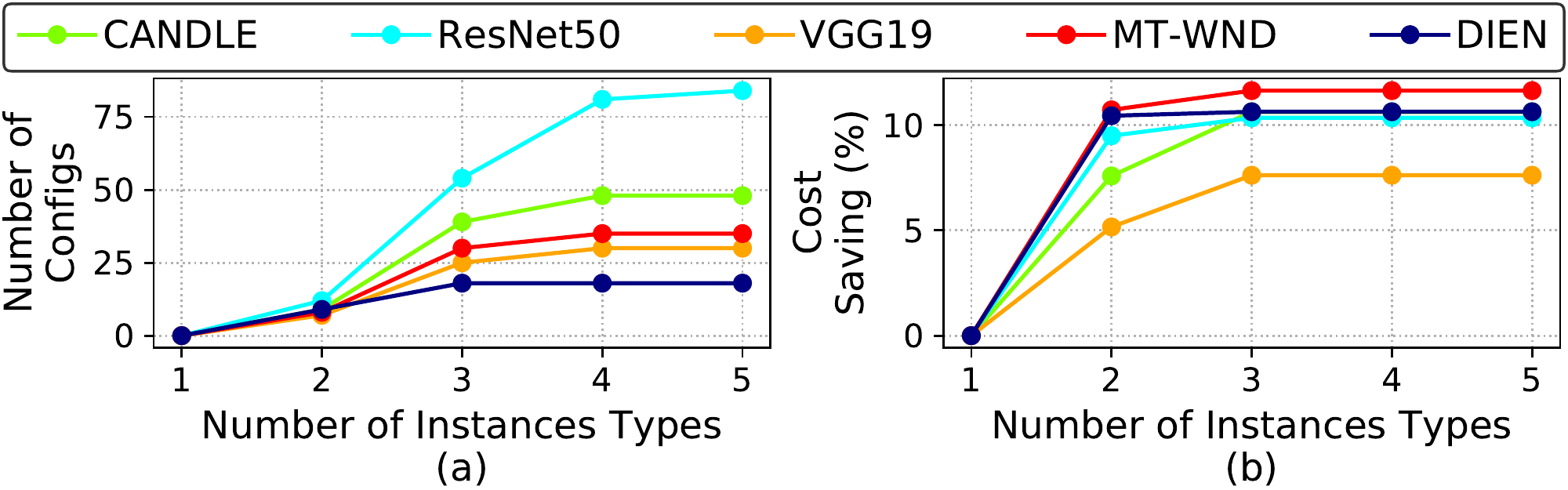}
    \hrule
    \vspace{-3mm}
    \caption{(a) The number of heterogeneous configurations better than the best homogeneous configuration starts saturating beyond three unique instance types in a heterogeneous configuration. (b) The cost savings from heterogeneous settings saturates after the number of unique instance types in a heterogeneous configuration goes beyond three.}
    \label{fig:metho_2}
\end{figure}

% \begin{figure}[t]
%     \centering
%     \subfloat[]{\includegraphics[scale=0.5]{figures/eval_1_het.pdf}}
%     \hspace{0.4cm}
%     \subfloat[]{\includegraphics[scale=0.5]{figures/eval_1_het_prime.pdf}}
%     \vspace{0.1cm}
%     \hrule
%     \vspace{-0.3cm}
%     \caption{(a) Optimal heterogeneous configurations reduce cost significantly over optimal homogeneous configurations across all models. (b) This cost improvement is achievable even using the same diverse pool of instances for all models.}
%     \label{fig:eval_1}
%     \vspace{-0.5cm}
% \end{figure}

%\subsection*{Saving Cost with Heterogeneous Instances}\vspace{0.2cm}

\subsection{Cost Savings with \sol{}} \vspace{1mm} %\vspace{2mm}

To demonstrate the cost savings of diverse pool configuration of \sol{}, we compare the diverse pool against the homogeneous instance it is based on. The pool composition is shown in Table~\ref{table:3}. We find that the effective diverse pool for models of the same category (general DNN/CNN or recommendation) tend to be common. As in Table~\ref{table:3}, the effective diverse pool is the same for CANDLE, ResNet50 and VGG19 models, and the same pool is used for MT-WND and DIEN models. Besides the two recommendation models in the table, we also tested on various other recommendation models in~\cite{gupta2020architectural,he2017neural,cheng2016wide}. The diverse pool (\texttt{g4dn,c5,r5n}) yields similar cost saving for the other models as well (results not shown for brevity). This is because different models of the same category tend to have similar computation characteristics, thus the performance of all the instances have similar ranks even across different models. 

Users who are aware of their model architecture can take advantage of this feature to quickly form a diverse pool and explore for a lower cost configuration without extensive offline knowledge about the instance performances. The cardinality of the pool is configured to be three because our empirical result (Fig.~\ref{fig:metho_2}) shows that using more than three diverse types slows down on generating new configurations that outperform the baseline (opportunity land for \sol{}) and the top cost savings stop increasing (benefits of \sol{}) -- this works in the favor of \sol{} since it reduces the search space.\vspace{0.2cm}

% \begin{figure}[t]
%     \centering
%     %\subfloat[][]{\includegraphics[scale=0.47]{figures/log_normal.pdf}}\hfill
%     \subfloat[][]{\includegraphics[scale=0.67]{figures/eval_1_het.pdf}}
%     \hfill
%     \vspace{0.1cm}
%     \hrule
%     \vspace{-0.3cm}
%     \caption{\revision{(a) Histogram of batch size samples and a log-normal probability density function (PDF) as comparison.} (b) Optimal heterogeneous configurations reduce cost significantly over optimal homogeneous configurations across all models.}
%     \label{fig:eval_1}
%     \vspace{-0.4cm}
% \end{figure}

% \begin{figure}[t]
%     \centering
%     \subfloat[][]{\includegraphics[scale=0.47]{figures/eval_1_het.pdf}}\hfill
%     \subfloat[][]{\includegraphics[scale=0.47]{figures/eval_8_gaussian.pdf}}
%     \hfill
%     \vspace{0.1cm}
%     \hrule
%     \vspace{-0.3cm}
%     \caption{\todo{Option: Merging Log-norm and Gaussian. Remove this figure and keep 10, 11, or keep this figure and remove 10, 11.}}
%     \label{fig:eval_1_merged}
%     \vspace{-0.4cm}
% \end{figure}

\begin{figure}[t]
    \centering
    \includegraphics[scale=0.48]{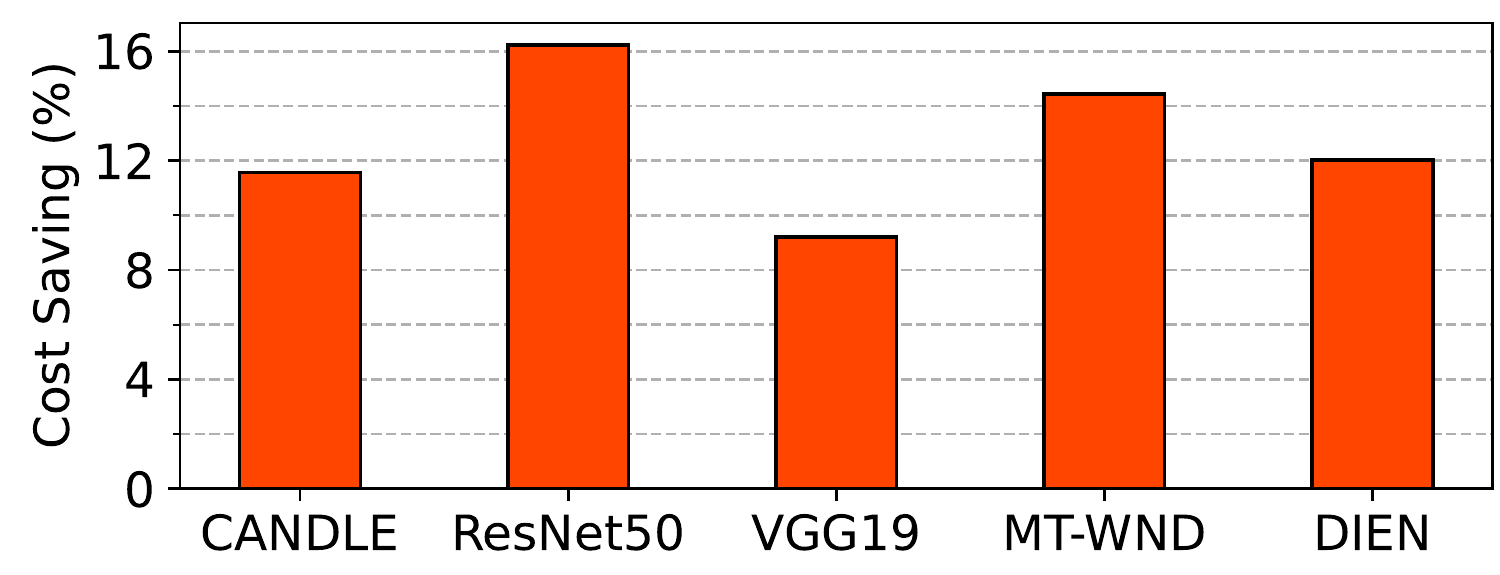}
    \vspace{2mm}
    \hrule
    \vspace{-3mm}
    \caption{Optimal heterogeneous configurations reduce cost significantly over optimal homogeneous configurations across all models.}
    \label{fig:eval_1_log_normal}
\end{figure}

\begin{figure*}[t]
    \centering
    \includegraphics[scale=0.49]{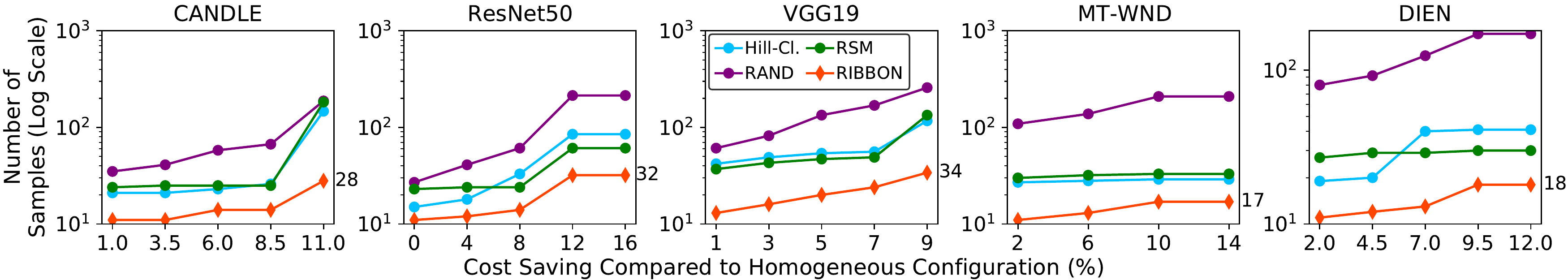}\newline
    %\vspace{0.1cm}
    \hrule
    \vspace{-3mm}
    \caption{\sol{} requires the fewest number of configuration samples before reaching different cost saving targets across all models.}
    \label{fig:eval_2}
\end{figure*}

\begin{figure}[t]
    \centering
    \includegraphics[scale=0.47]{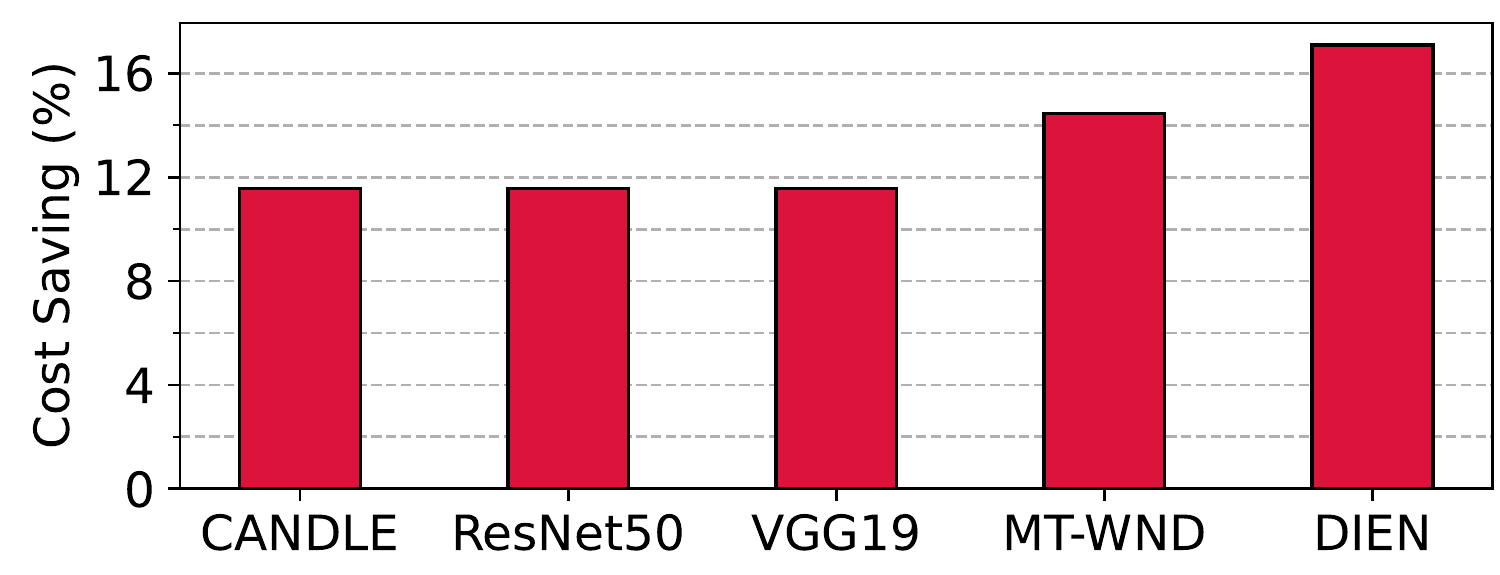}
     \vspace{2mm}
    \hrule
    \vspace{-3mm}
    \caption{\revision{\sol{} continues to provide significant cost savings over optimal homogeneous configurations even when the batch size follows a Gaussian distribution.}}
    \label{fig:eval_1_gaussian}
\end{figure}

\noindent\textbf{Using a heterogeneous instance configuration can significantly save cost while meeting the QoS target.} In Fig.~\ref{fig:eval_1_log_normal}, when using a 3-type heterogeneous instance configuration, the minimum cost to meet the QoS is lower for all models (the results show improvement over the best homogeneous configuration instance). The cost savings range from 9\% (VGG19) to 16\% (ResNet50).

We found that such improvements do not tend to require specifically selected instance pools as they are model-specific. In fact, similar improvements are observed for other diverse instance pools (results not shown for brevity). Further, we note that even a single-digit percent cost saving can be significant when accumulated over time. A cost improvement of a few percent would mean millions of dollars of savings in expenditure. \revision {Our results (Fig.~\ref{fig:eval_1_gaussian}) also show that \sol's cost savings are not sensitive to the batch size distribution. It still provides significant cost savings even when the batch size follows a Gaussian distribution. This is primarily because of the flexibility a diverse instance pool provides. Varying the batch distribution would make particular diverse pool configurations ineffective anymore (i.e., higher cost, QoS violation), but would also make some other configurations more effective (i.e., lower cost within QoS). The \sol{} optimization framework does not make assumptions about batch size distribution. It is designed to explore the heterogeneous configuration space carefully in order to achieve the best savings fast, regardless of where the actual optimal configuration lies.}  

Next, we look at how \sol{} helps find the optimal configurations most effectively when compared to other competing techniques for finding \sol{}-like configuration. 

\subsection{Fast Convergence to the Optimal Diverse Pool Configuration with \sol{}}\vspace{1mm}%\vspace{0.2cm}

\noindent\textbf{Competing strategies.} To evaluate the efficiency of \sol{} in finding the optimal configuration, we compare it against multiple competing online schemes that search for optimal heterogeneous pool configurations.\newline 

\noindent\textbf{Random (RANDOM).} This is a relatively simple strategy that evaluates different random configurations in the search space. To make it more intelligent, we do not evaluate a randomly picked configuration if a previous configuration with a higher number of instances for each type does not meet the QoS target, or a previous configuration with a lower number of instances for each type meets the QoS at a lower cost. \newline 

\noindent\textbf{Hill Climbing (Hill-Climb).} This scheme is designed on the hill-climbing optimization principles, widely-used for fast search space exploration~\cite{gupta2020deeprecsys,chen2019parties,yildiz2009effective,shehab2017hybridizing}. We have customized and optimized for our use case by intelligently increasing and decreasing the number of instances based on the observed QoS and cost in a multi-dimensional search space. \newline 

\noindent\textbf{Response Surface Methodology (RSM).} The Response Surface Methodology (RSM)~\cite{bradley2007response} is an advanced technique used for exploring a certain unknown surface with a fixed number of samples. We employ an optimized 3-level 3-factor central composite face-centered design to explore the search space, since the central composite design is a widely used and one of the most promising RSM designs~\cite{ahmadi2005application,bashiri2011tuning,rajmohan2013application}. The RSM sampled configurations will be evaluated, and the scheme starts exploring around the most promising point.

%% file: sections/evaluation.tex
\begin{figure}[t]
    \centering
    \includegraphics[scale=0.55]{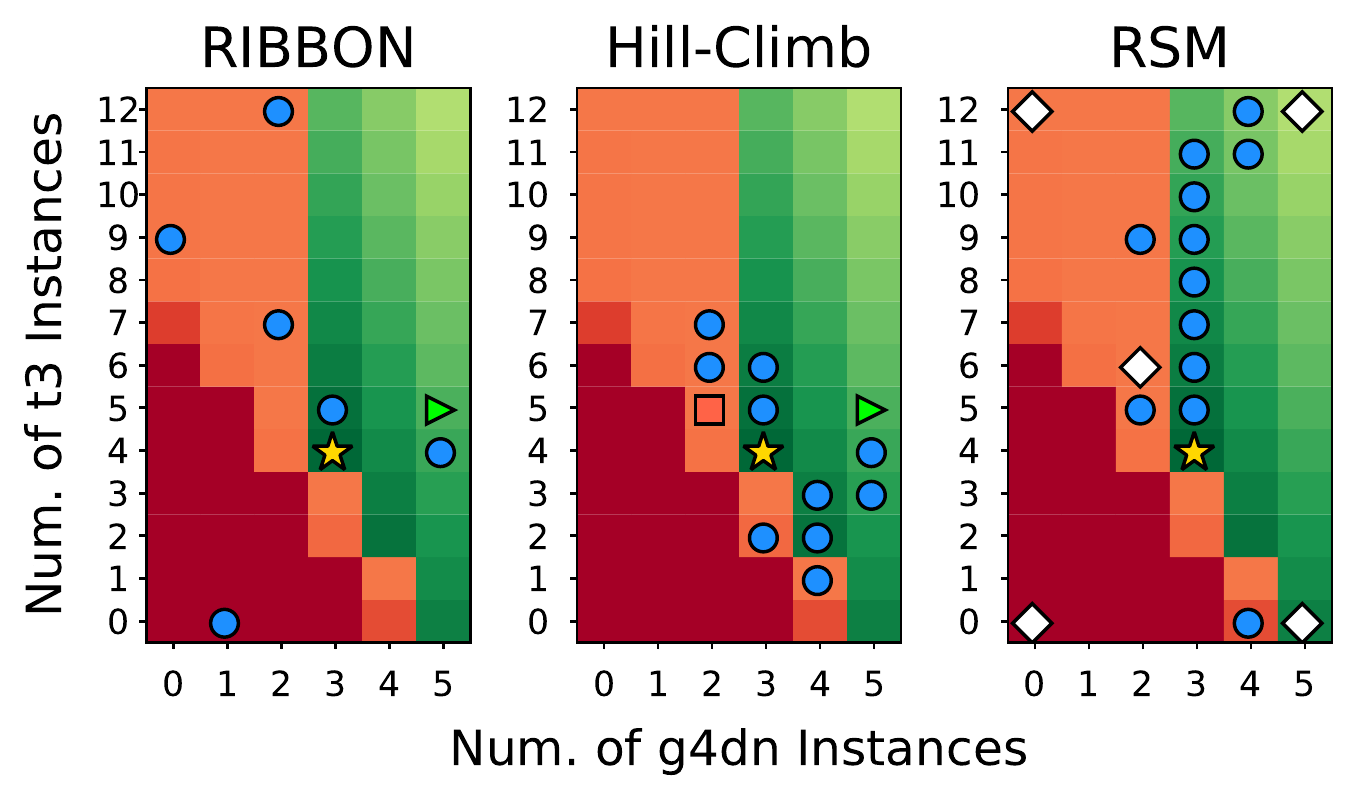}
    \hrule
    \vspace{-3mm}
    \caption{Evaluation samples of search method on a two-dimension MT-WND model example. Red grid: QoS violation configurations (lighter the better). Green grid: QoS satisfaction configurations (darker the better). \revision{Light green triangle} (5, 5): \sol{} and \hill{} starting point. \revision{Gold star} (3, 4): optimal configuration. \revision{Blue circles:} Explored configurations. \revision{Dark orange square}: \hill{} restart point. \revision{White diamonds}: \rsm{} central composite design samples.}
    \label{fig:eval_5}
\end{figure}

\noindent\newline\textbf{\sol{} converges to cost-saving heterogeneous configurations with the least number of samples compared to all other competing strategies.} In Fig.~\ref{fig:eval_2}, the results show the number of samples required to reach a certain cost-saving given the specific instance diverse pool for each model. In order to reach a higher cost saving compared to the homogeneous configuration, more configuration samples are needed to be evaluated. To find the maximum possible cost-saving, \sol{} requires evaluating less than 40 configurations out of roughly 1000s of possible configurations in total, and for the MT-WND and DIEN models, only about 20 evaluations are needed. Compared to \sol{}, the other competing strategies require an order of magnitude more number of samples for the CANDLE model, and more than 2 to 3 times more samples for ResNet50, VGG19, MT-WND, and DIEN models. 

The performance of \hill{}, \random{} and \rsm{} varies across models as \rsm{} is better than \hill{} on ResNet50 and DIEN models while being close to each other on CANDLE, VGG19, and MT-WND models, meaning the central composite design sampled configurations are closer to the optimal configuration for ResNet50 and DIEN models; \random{} consistently requires a lot more samples to reach optimum across most models, but \rsm{} and \hill{} requires the same number of samples as \random{} to reach the optimum for CANDLE, because in this case, the search space has a large number of local optima. Overall, \sol{} always consistently outperforms all other strategies to reach either an intermediate cost-saving configuration or the optimal configuration.\newline %These trends also hold true when the same diverse pool is used for all models, \sol{} still outperforms all other strategies regardless of the pool choice (the result is not shown for brevity and due to similarity to the results in Fig.~\ref{fig:eval_2}). 

\textit{To dive further into why \sol{} performs better than other intelligent techniques like \hill{} and \rsm{}, Fig.~\ref{fig:eval_5} shows a simplified two-dimensional example of the configurations explored by these techniques on the MT-WND model}. The two instance types in the diverse pool are \texttt{g4dn} and \texttt{t3}. \sol{} actively searches for configurations across the whole space based on its best estimation from the surrogate model. It sampled the dark red region but was able to avoid extra samples nearby, and it was able to find the optimal configuration in eight evaluations. 

On the other hand, \hill{} initially climbed to a local optimum at point (4,3), sampled every possible point with a cheaper cost, then realized there is no way to go as both (3,2) and (4,1) violate QoS. It then randomly sampled a new point (dark orange) to restart, and successfully climbed to the global optimum in 13 evaluations in total. Lastly, \rsm{} first sampled the five fixed points marked in white, and then, it started searching near the best one (5,0). However, (5,0) is already a local optimum, so it switched to (5,12) and found the global optimum from there (18 evaluations in total). This shows that the \rsm{} sampling does not fit the search space well as the sample (2,6) is actually the closest to the optimal configuration in distance, but was not picked as a starting point due to its QoS violation. In both \hill{} and \rsm{}, many configurations around a local optimum are evaluated before moving on, but in \sol{}, being stuck at a local optimum is not observed due to the exploration aspect of its BO engine. \sol{}'s active pruning helps it to continually identify and avoid the set of sub-optimal configurations, thus making its search process faster. 

\begin{figure}[t]
    \centering
    \includegraphics[scale=0.47]{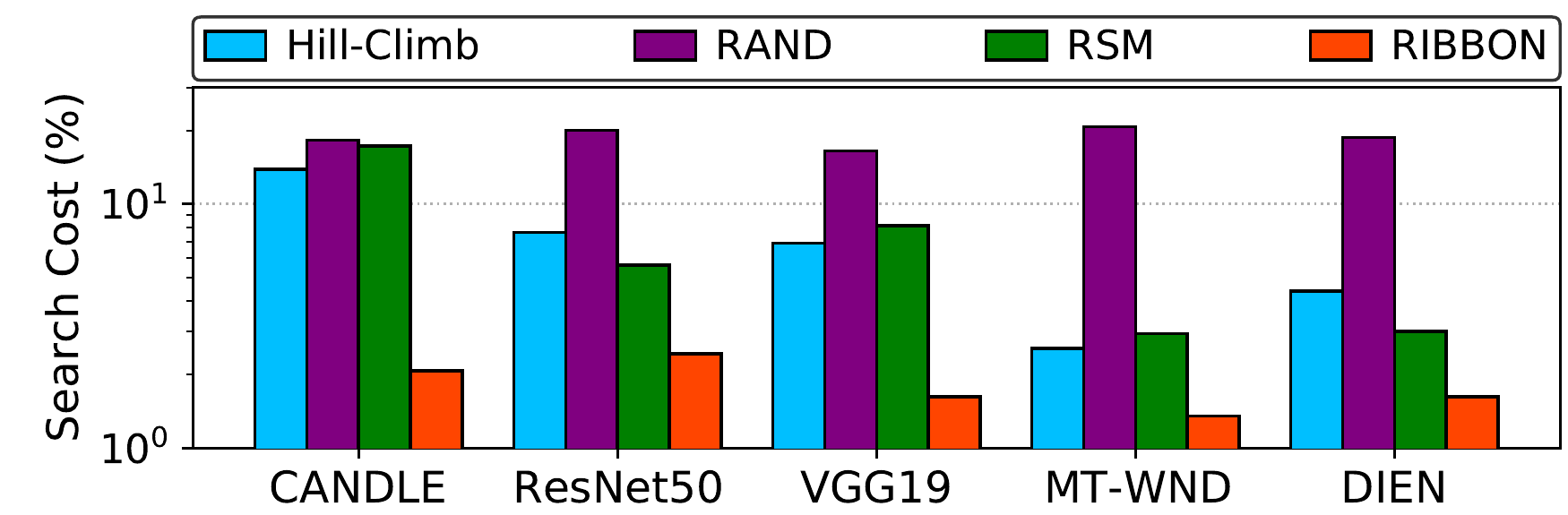}
    \vspace{2mm}
    \hrule
    \vspace{-3mm}
    \caption{Exploration cost (as a percentage of exhaustively sampling every configuration) of finding the optimal configuration is the lowest with \sol{}, as compared to competing techniques.}
    \label{fig:eval_6}
\end{figure}

\subsection{Low Exploration Cost and High Benefits with Relaxed QoS Constraints}\vspace{1mm}

\noindent\textbf{\sol{}'s total cost of exploration evaluation is also the lowest among all strategies. It also explores QoS-violating configurations the least number of times.} While \sol{} minimizes the number of configurations sampled during exploration, a desirable goal is to also minimize the total cost of exploration -- although we note that running the optimal diverse pool configuration for some time after it has been achieved by any method would outweigh the cost during exploration. Nevertheless, the cost can be further lowered by exploring fewer and cheaper configurations. Fig.~\ref{fig:eval_6} shows the total cost to find the optimal configuration as a percentage of the cost for evaluating all available configurations exhaustively (exhaustive-search cost). For \sol{}, this exploration cost is less than 3\% for all models, while for other strategies it would cost 10 to 20\% of the exhaustive-search cost to find the same optimal configuration as \sol{}. 

\begin{figure}[t]
    \centering
    \includegraphics[scale=0.46]{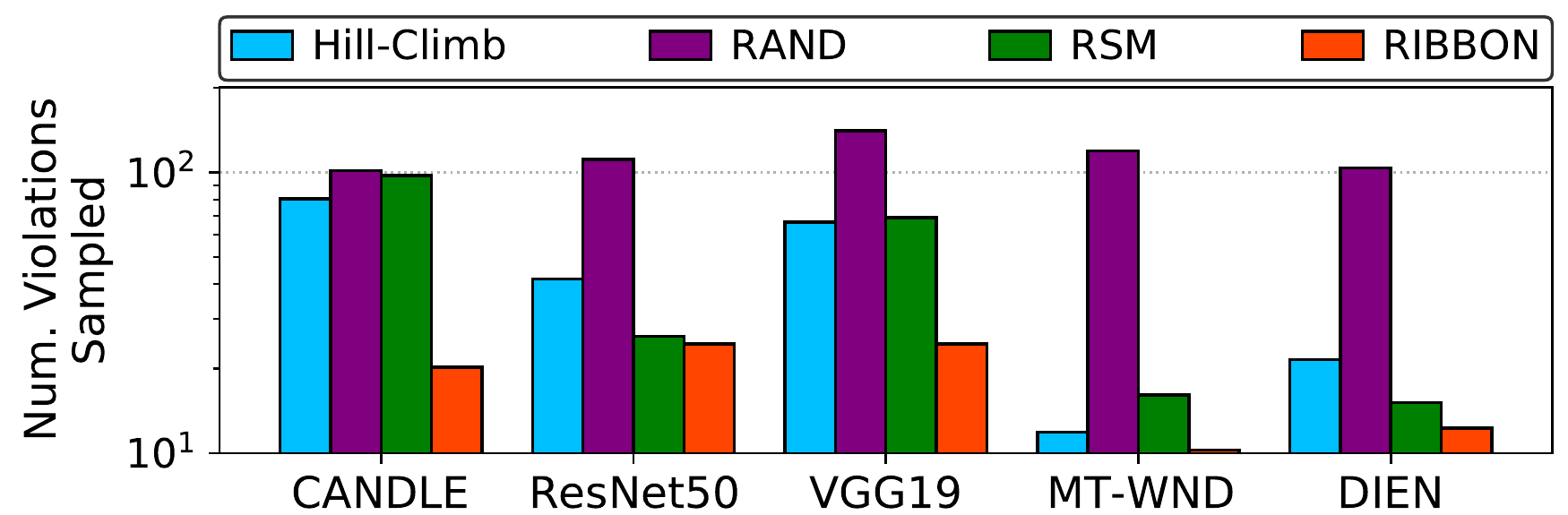}
    \vspace{2mm}
    \hrule
    \vspace{-3mm}
    \caption{Number of QoS violating configurations sampled before finding the optimal configuration is the lowest with \sol{}.}
    \label{fig:eval_4}
\end{figure}

Another desirable characteristic during the exploration process is to minimize the sampling number of QoS violating configurations. During the search process, it is unavoidable to evaluate configurations that violate QoS. Even though it is a temporary process, inference service providers would still find it sub-optimal to sample configurations that violate the QoS. In Fig.~\ref{fig:eval_4}, during the search process before finding an optimal configuration, the total number of evaluated QoS-violating configurations are compared among the competing techniques. Because \sol{} requires much fewer samples, it gets an advantage on limiting the number of QoS-violating samples. In fact, for models CANDLE, VGG19, MT-WND, and DIEN, \sol{} samples the least number of QoS-violating configurations during the exploration phase. For example, for model CANDLE, \sol{} samples 20 QoS-violating configurations, while the other three techniques sample up to 100 QoS-violating configurations. Interestingly for model ResNet50, \rsm{} is very close to \sol{}. In this case, the optimal configuration is close to the central composite design sampled configurations. So, \rsm{} is able to quickly converge to it from one of its initial samples. 

\revision{These results also support that spawning \sol{} exploration is always recommended due to its low exploration cost (less than 3\% of exhaustive search, to be recouped within a few hundred queries when steady-state is reached) and lower number of QoS violations even during the exploration. Running \sol{} provides the opportunity to avail both transient and steady-state savings.}

\begin{figure}[t]
    \centering
    \includegraphics[scale=0.46]{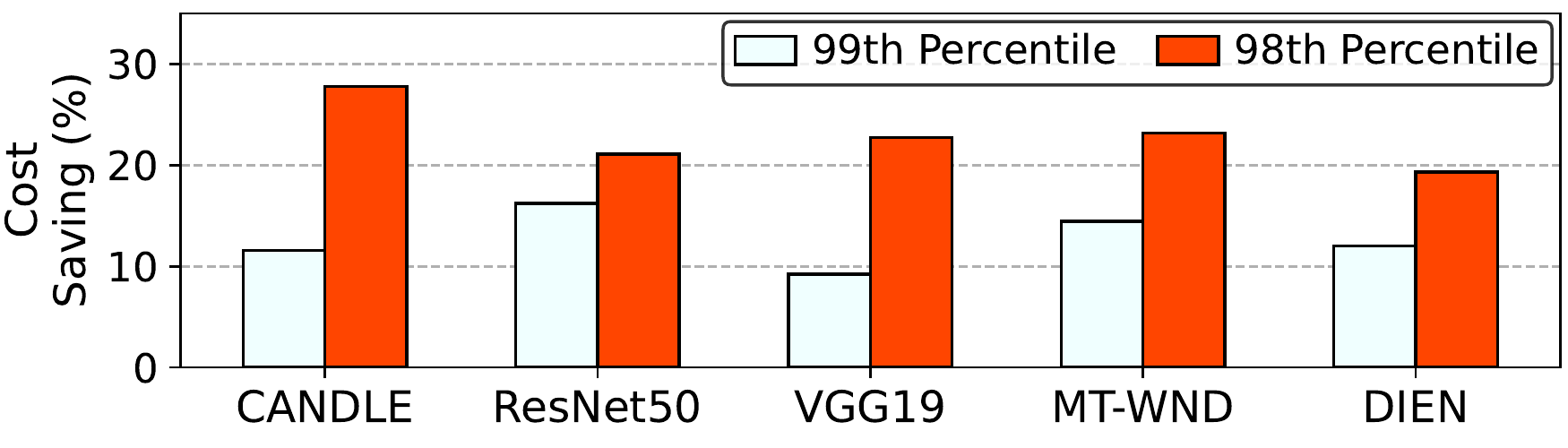}
    \vspace{2mm}
    \hrule
    \vspace{-3mm}
    \caption{\sol{} is able to improve the cost savings even further when the QoS target is relaxed to the 98$^{th}$ percentile tail latency.}
    \label{fig:eval_7}
\end{figure}

\begin{figure*}[t]
    \centering
    \includegraphics[scale=0.49]{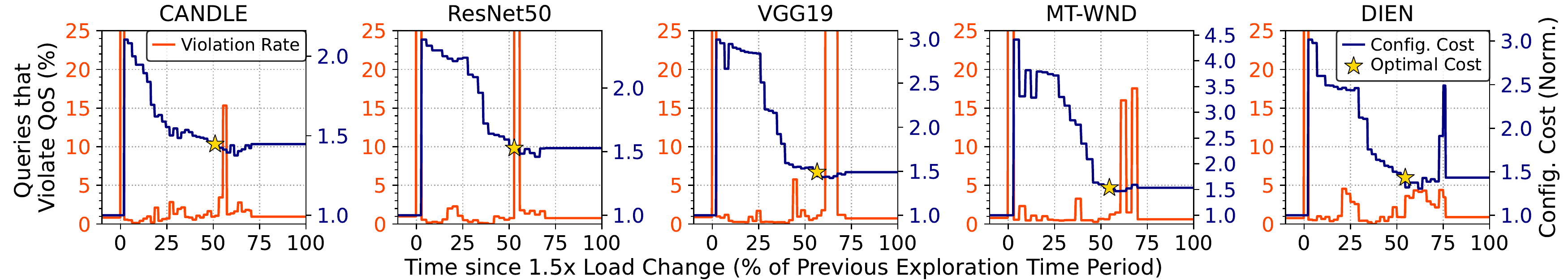}
    \vspace{2mm}
    \hrule
    \vspace{-3mm}
    \caption{\sol{} quickly responds to load change with new instance configurations and starts searching for new optimal cost configuration.}
    \label{fig:eval_8}
\end{figure*}

Our results so far have focused on strict $99^{th}$ percentile tail latency targets as default for all models. To better understand the effectiveness of \sol{}, we relax the QoS target since some service providers can use relaxed QoS targets for certain types of customers. We note making the QoS-target more strict (e.g., $99.9^{th}$ percentile tail latency) was not possible for some models since even the most performant AWS instances are not able to reach this target.\vspace{0.2cm}

\noindent\textbf{The benefit of using a diverse pool is more significant when the QoS is relaxed.} Fig.~\ref{fig:eval_7} shows that if the QoS is relaxed slightly to the $98^{th}$ percentile tail latency (p98), the cost savings compared to homogeneous configurations become more significant. This is because a more relaxed QoS gives more freedom to use the lower performance but also lower cost instances -- this opportunity is exploited by \sol{} which builds a diverse pool, but remains unexploited by homogeneous pool instances. For example, the p98 optimal configuration for the CANDLE model consists of 2 \texttt{c5a}, 1 \texttt{m5} and 6 \texttt{t3} instances. This configuration is 17\% less expensive than the optimal configuration to achieve the $99^{th}$ percentile tail latency (p99) QoS (6 \texttt{c5a} and 2 \texttt{t3} instances). Thus, \sol{} is capable of achieving further gains for workloads with lower QoS requirements. Nevertheless, we chose stricter QoS requirements as default to present conservative results. 

\subsection{\sol{}'s Quick and QoS-friendly Adaption to Load Fluctuations }\vspace{1mm}

\noindent\textbf{When the load level changes, \sol{} can quickly adjust to configurations that can serve the new load better, and start searching for the optimal configuration for the new load.} In Fig.~\ref{fig:eval_8}, \sol{}'s adjustment to load scale is demonstrated as time series. For each model, the x-axis represents the time as a percentage of total time to find the optimal configuration prior to the load scaling. For example, suppose it takes the CANDLE model 10 minutes to reach an optimal configuration from scratch, then after some time, the load becomes 1.5 times heavier, and a 50\% time means \sol{} spends 5 minutes to reach the new optimal configuration on the new load. The load change occurs at time 0. The orange curve shows the percentage of queries that violate QoS for each configuration \sol{} explores. It should stay within 1\% for the service $99^{th}$ tail latency target to be satisfied. The blue curve shows the cost (\$) of each configuration normalized to the optimal configuration cost before the load changes. \sol{}'s goal is to find a point such that the orange curve is below 1\% while the blue curve is the lowest. We have several observations from this figure. 

First, \sol{} can quickly respond to the load change by adjusting to a more expensive and better performance configuration so that the QoS violations are reduced significantly. The queries that violate QoS experience a surge on the load change, reaching more than 95\% violation (not shown in Fig.~\ref{fig:eval_8} as lower violation regions are more interesting). This is because the current configuration's allowable throughput is saturated, most of the queries have to stay in the queue waiting, the accumulated queries make the queue longer and longer, resulting in violation for almost all queries. Once \sol{} detects such behavior, it resets the search space target values and based on prior knowledge, provides estimation to previously evaluated configurations on the new load, prunes away configurations that would definitely violate QoS. This process greatly helps \sol{} avoid the QoS violation region, explore the QoS satisfaction region and maneuver the diverse pool towards a lower-cost configuration. In the end, \sol{} is able to find an optimal configuration that is also around 1.5 times more expensive than the previous cost. Notice that the system does not know how much the load has changed when we apply the load change, so \sol{} does not know the new optimal cost is around 1.5 times more expensive prior to exploration.

Second, we note that once \sol{} has established an optimal configuration at a given load, the time it takes to converge to a new load optimum tends to be shorter than the previous convergence time. In Fig.~\ref{fig:eval_8}, the time to reach the optimal cost star is less than 60\% of the time needed to reach the previous optimum. This is also due to the benefit of estimation and pruning, as some configurations are considered as ``already explored'' before the exploration for new load actually starts, avoiding unnecessary sampling of these configurations that would greatly violate the QoS and point \sol{} to more promising regions. It is also worth noticing that after the optimal configuration is found, \sol{} tends to explore some configurations that would greatly violate the QoS again (red spikes after the star in Fig.~\ref{fig:eval_8}). This is because \sol{} has finished exploiting the more promising configurations, and starts to sample the less promising but unexplored configurations. This is also a good sign that \sol{} should stop searching and settle on the best configuration until this point. In practice, it is possible to reduce the violation spikes during exploration by monitoring the query queue. If the queue stacks beyond a limit, it means the configuration may violate QoS. The evaluation of this configuration can be terminated early to avoid more queries from violating the QoS. 

% Overall, \sol{} is effective in multiple desirable aspects and yields significant monetary benefits that aggregates over time and resources. 

%------ SCRATCH SPACE-------------------

%only \$1.88/hr in total. Thus, \sol{} is capable of achieve further gains for workloads with lower QoS requirements.
%For example, the optimal configuration for the RM2 model to achieve the $99^{th}$ percentile tail latency (p99) QoS consists of 6 \texttt{c5a} and 3 \texttt{r5} instances, which costs \$2.23/hr in total. But, the 

% \begin{figure}[H]
%     \centering
%     \includegraphics[scale=0.12]{figures/eval_3_breakdown.jpg}
%     \vspace{0.1cm}
%     \hrule
%     \vspace{-0.3cm}
%     \caption{}
%     \label{fig:eval_3}
%     \vspace{-0.5cm}
% \end{figure}

%Notice that the difference between \sol{} and its competing strategies enlarges as the cost saving target gets closer to optimal, for example when the cost saving is more than 7\% on RM2 and 9.5\% on DIEN. This indicates that the 

%However, reducing the number of \texttt{c5a} by one in the optimal heterogeneous configuration at p99 QoS would drastically drop the performance, which cannot meet the p98 target (Sec.~\ref{sec:motiv} Fig.~\ref{fig:char_7}).

%% file: sections/conclusion.tex
\section{Conclusion}
\label{sec:concl}

To the best of our knowledge, \sol{} is the first work to employ the idea of using a heterogeneous pool of cloud computing instances to significantly improve the cost-effectiveness and performance-efficiency of deep learning model inferences. \sol{} finds the most-optimal instance configuration to jointly optimize two competing goals: meeting the QoS and minimizing cost. To solve this problem, \sol{} uses a BO-based exploration-exploitation approach to find the optimal instance configuration. Our evaluation confirms that \sol{} can yield up to 16\% cost savings without extensive offline knowledge and can adapt to load changes quickly.

%\sol{}'s BO-based solution has the ability to quickly adapt to the changes in application and hardware and performs several optimizations to speed up its search process.  